\documentclass[%
 reprint,
 amsmath,amssymb,
pra,
]{revtex4-2}

\usepackage{graphicx}
\usepackage{dcolumn}
\usepackage{multirow}
\usepackage{bm}

\usepackage{hyperref}
\usepackage{qcircuit}
\usepackage{graphicx}
\graphicspath{{figures/}}
\usepackage{dcolumn}
\usepackage{bm}
\usepackage{bbold}
\usepackage[utf8]{inputenc}
\usepackage[T1]{fontenc}
\usepackage{mathptmx}
\usepackage{mathtools}
\usepackage{xcolor}
\usepackage{tikz}
\usepackage{braket}
\usepackage{wrapfig}

\definecolor{ao(english)}{rgb}{0.0, 0.5, 0.0}

\begin{document}

\preprint{AIP/123-QED}

\title[YB-QC]{Quantum time dynamics of 1D-Heisenberg models employing the Yang-Baxter equation for circuit compression}

\author{Sahil Gulania}
\email{sgulania@anl.gov}
\affiliation{Mathematics and Computer Science Division, Argonne National Laboratory, Lemont, Illinois 60439, United States}

\author{Bo Peng}
\email{peng398@pnnl.gov}
\affiliation{Physical and Computational Sciences Directorate, Pacific Northwest National Laboratory, Richland, Washington 99352, United States}

\author{Yuri Alexeev}
\email{yuri@anl.gov}
\affiliation{Computational Science Division, Argonne National Laboratory, Lemont, Illinois 60439, United States}

\author{Niranjan Govind}
\email{niri.govind@pnnl.gov}
\affiliation{Physical and Computational Sciences Directorate, Pacific Northwest National Laboratory, Richland, Washington 99352, United States[\phantom]}

\date{\today}
\begin{abstract}
Quantum time dynamics (QTD) is considered a promising problem for quantum supremacy on near-term quantum computers. However, QTD quantum circuits grow with increasing time simulations. This study focuses on simulating the time dynamics of 1-D integrable spin chains with nearest neighbor interactions. We show how the quantum Yang-Baxter equation can be exploited to compress and produce a shallow quantum circuit. With this compression scheme, the depth of the quantum circuit becomes independent of step size and only depends on the number of spins. We show that the compressed circuit scales quadratically with system size, which allows for the simulations of time dynamics of very large 1-D spin chains. We derive the compressed circuit representations for different special cases of the Heisenberg Hamiltonian. We compare and demonstrate the effectiveness of this approach by performing simulations on quantum computers.

\end{abstract}
\maketitle

\section{Introduction}


Simulation of statistical mechanical models is a vital application for classical and quantum computing \cite{alexeev2021}. It is also well known that the partition function of a $d+1$-dimensional classical system can be mapped to the partition function of a $d$-dimensional quantum system \cite{suzuki1976relationship, sachdev2011quantum, shankar2017quantum}. This deep classical to quantum connection can give insights into quantum universality classes using appropriate classical counterparts. It might allow one to use these effective models to study critical points and phase transitions of magnetic systems, where the spins of the magnetic systems are treated quantum mechanically. 

Quantum Ising and Heisenberg models \cite{sutherland2004beautiful} represent some of the simplest models that can describe the behavior of magnetic systems. However, 2D and 3D-quantum lattice simulations remain challenging for classical computing. A possible solution is to use quantum computing, given that it is only natural to simulate quantum systems with quantum computers as suggested by Benioff \cite{benioff1980computer} and Feynman \cite{feynman1982simulating}. The idea of using quantum computing on quantum devices is compelling since it paves the way for systematic improvements of quantum technologies. The exploration of this approach is at the core of several research efforts in quantum information sciences \cite{DOE-QIS}. In particular, the development of new generations of quantum models and associated simulations on existing and upcoming Noisy Intermediate-Scale Quantum (NISQ)  \cite{preskill2018quantum} devices is of particular interest.

Quantum integrable systems \cite{jimbo1989ybe} typically refer to systems where the dynamics are two-body reducible. Put another way, even though the Hilbert space increases exponentially with increasing system size in these systems, the two-body reducibility, in combination with the algebraic Bethe ansatz,  \cite{bethe1931theorie,batchelor2007bethe,sutherland2004beautiful} can be used to obtain explicit solutions, under certain conditions, by solving a set of non-linear equations that scales only linearly with system size. The quantum Yang–Baxter equation or star-triangle relation \cite{yang1967some, baxter2016exactly, jimbo1989ybe, perk2006encyclopedia} is a consequence of this factorization. The algebraic formulation of quantum integrable systems makes them ideal tools to study a broad range of low dimension physical models. Historically, the isotropic interacting quantum spin chain, or Heisenberg model  \cite{heisenberg1928theorie} was the first quantum integrable system, whose exact eigenstates were obtained through the Bethe ansatz approach as a superposition of plane-waves  \cite{bethe1931theorie,takahashi1971one,karbach1997introI,karbach1998introII,karbach1998introIII,faddeev1996how}. Other quantum integrable models include the Lieb–Liniger model  \cite{lieb1963exactI,lieb1963exactII}, the Hubbard model  \cite{essler2005one}, Calogero-Sutherland model  \cite{calogero1969ground,calogero1969solution,calogero1971solution,sutherland1971quantumI,sutherland1971quantumII,sutherland2004beautiful}, models from quantum field theory like the sine-Gordon model  \cite{johnson1973vertical,luther1976eigenvalue}, and several sub-classes of the Heisenberg model (e.g. XXZ model)  \cite{heisenberg1928theorie,orbach1958linear}.

An important yet open question in quantum integrable systems is time evolution, or the response dynamics where the system responds to a change of parameters. This requires sufficiently accurate control of the time evolution of the system which is governed by the time-dependent Schr\"{o}dinger or Dirac equation of the quantum state in the Hilbert space. The time evolution problem is also closely related to the computation of the asymptotic state of quantum integrable models, or in the more general sense the thermalization and ergodic/non-ergodic behaviors of these models. Several conjectures have been proposed in this context  \cite{rigol2007relaxation,cassidy2011generalized,caux2013time}. For example, in the Heisenberg spin chain, quantum quenches of the XXZ model has been studied by embedding the generalized Gibbs ensemble hypothesis into the quantum transfer matrix framework  \cite{fagotti2013dynamical,pozsgay2013generalized}. Nevertheless, these studies are still far from conclusive.  Other types of quantum integrability are known in explicitly time-dependent quantum problems, such as the driven Tavis-Cummings model \cite{sinitsyn2016solvable, sinitsyn2018integrable}.

It is well known that quantum circuits representing quantum time dynamics (QTD) grow with increasing time simulations. In the present era of noisy quantum computers, it has become a necessity for circuits to be as shallow as possible for meaningful results. With this as the overarching theme, we focus on QTD of 1-D integrable spin chains with nearest neighbor interactions in this paper. We show how the quantum Yang-Baxter equation can be used to compress and produce a shallow quantum circuit, where the depth becomes independent of step size and only depends on the number of spins. The compressed circuit scales quadratically with system size, which allows for the simulations of time dynamics of very large 1-D spin chains. Compressed circuit representations are derived for different special cases of the Heisenberg Hamiltonian. As a proof of principle, we demonstrate the effectiveness of this approach by performing simulations of the Heisenberg XY model (or XY model in brief) on quantum devices. The time evolution of the XY model is an active area of research and has been approached from many directions. Verstraete and co-workers~\cite{PhysRevA.79.032316} utilized the quantum Fourier transform with the Bogoliubov transformation to perform time dynamics efficiently on a quantum computer. Recently, in a similar study to this work, Bassman and co-workers~\cite{bassman2021constantdepth} also reported simulations of the XY model by conjecturing the relationship between the reflection symmetry and YBE and provided numerical evidence for the YBE transformation. In this paper we go further by putting the relationship on a firm footing and derive the analytical expressions. For the rest of this paper we will tacitly assume the quantum Yang-Baxter equation (YBE) and omit quantum for brevity. 

\section{Theory}
\label{theory}
For completeness, we start by reviewing relevant background material, with a brief introduction to the Heisenberg Hamiltonian, quantum time dynamics, and the Yang-Baxter equation.

\subsection{Heisenberg Hamiltonian}
The Heisenberg Hamiltonian \cite{fazekas1999lecture, skomski2008simple, pires2021theoretical} is widely used to study magnetic systems, where the magnetic spins are treated quantum mechanically. The Hamiltonian, including only spin-spin interactions, can be written as
\begin{equation}
    \hat{H} = 
    -\sum_{\alpha}\{J_{\alpha}\sum_{i=1}^{N-1} \sigma_{i}^{\alpha}\otimes \sigma_{i+1}^{\alpha}\} 
\end{equation}
where, $\alpha$ sums over $\{x,y,z\}$, the coupling parameters $J_{\alpha}$ denotes the exchange interaction between nearest neighbour spins along the $\alpha-$direction, $\sigma^{\alpha}
_{i}$ is the $\alpha$-Pauli operator on the $i$-th
spin. Interaction with the magnetic field can be included in the above Hamiltonian as 
\begin{equation}
    \hat{H}_{in}(t) = \hat{H}- h_{\beta}(t) \sum_{i=1}^{N} \sigma_{i}^{\beta}
\end{equation}
where, $h_{\beta}(t)$ is the time amplitude of the external magnetic field along the $\beta\in\{x,y,z\}$ direction. 
Several variations of this model are known in the literature, which are categorized depending on the relation between $J_x$, $J_y$, and $J_z$. This Heisenberg Hamiltonian represents quantities based on the electronic structure of the system, where the Coulomb interaction and hopping are mapped onto spin variables \cite{illas2006unified,david2017physical}.

A simple variant of the Heisenberg model is the one-dimensional XY model that was first introduced and solved by Lieb, Schultz, and Mattis  \cite{lieb1961two} in the absence of a magnetic field, and later by Katsura  \cite{katsura1962statistical,katsura1963statistical} and Niemeijer  \cite{niemeijer1967some} in a finite external field. The XY model describes a one-dimensional lattice with spin variables labeling every lattice point. The spins are limited to interact only with their nearest neighbors in an anisotropic way. As a quantum integrable model, all static correlation functions of the XY model can be computed in terms of Toeplitz determinants  \cite{mehta1989matrix}. For example, beside the fundamental correlation functions, one can also compute the Emptiness Formation Probability  \cite{shiroishi2001emptiness,abanov2003emptiness,franchini2005asymptotics}, the Von Neumann and Renyi entanglement entropies  \cite{jin2004quantum,its2005entanglement,its2006entropy,peschel2004on,franchini2007ellipses,franchini2007renyi}, and even some non-equilibrium properties  \cite{silva2008statistics,calabrese2011quantum,calabrese2012quantumI,calabrese2012quantumII,happola2012universality,bucciantini2014quantum,bayocboc2015exact} of the XY model. 

It has been shown  \cite{barouch1970statI,barouch1970statII,barouch1970statIII,mccoy1971statIV} that the excitations in the XY model are non-local free fermions that give rise to a non-trivial phase diagram at zero temperature. There exists two quantum phase transitions (QPTs) in this diagram featuring the universality of the anti-ferromagnetic Heisenberg chain (i.e. the isotropic XX model) and the quantum Ising model, respectively. In particular, the isotropic XX model corresponds to free fermions hopping on a lattice, and the Ising model corresponds to a transition from a doubly degenerate ground state to a single ground state  \cite{franchini2017introduction,mussardo2010statistical}.

\subsection{Time Evolution}
Quantum state evolution \cite{kosloff1988time,tannor2007introduction} is governed by the Schr$\ddot{o}$dinger or Dirac equation
\begin{equation}
    i\hbar\frac{\partial}{\partial t}\ket{\psi(t)} = \hat{H}\ket{\psi(t)}
\end{equation}
The solution to the above equation can be expressed as
\begin{align}
    \ket{\psi(t)} = e^{-i\hat{H}t/\hbar}\ket{\psi(0)}
\end{align}
where, $e^{-i\hat{H}t/\hbar}$ is the evolution operator. In the 1D-Heisenberg
model, with the exception of $N=2$, all the elements in the Hamiltonian do not commute with each other, and hence the exponential of $\hat{H}$ cannot be written as a product of exponentials. For $N=2$, 
\begin{equation}\label{eq:2spin}
    e^{-i\hat{H}t/\hbar} = \prod_{\alpha}
    e^{iJ_{\alpha}t(\sigma_{1}^{\alpha}\otimes\sigma_{2}^{\alpha})/\hbar}
\end{equation}
where each term is straightforward to evaluate as shown below 
\begin{equation}\label{eq:decomposition}
\begin{aligned}
   e^{iJ_{x}t(\sigma_{1}^{x}\otimes\sigma_{2}^{x})/\hbar} =&  
   \begin{pmatrix}
   \cos(\theta_{x}) & 0 & 0 & i\sin(\theta_{x}) \\
   0 & \cos(\theta_{x}) & i\sin(\theta_{x}) & 0 \\
   0 & i\sin(\theta_{x}) & \cos(\theta_{x}) & 0 \\
   i\sin(\theta_{x}) & 0 & 0 & \cos(\theta_{x}) 
   \end{pmatrix} \\ \\
   e^{iJ_{y}t(\sigma_{1}^{y}\otimes\sigma_{2}^{y})/\hbar} =&  
   \begin{pmatrix}
   \cos(\theta_{y}) & 0 & 0 & -i\sin(\theta_{y}) \\
   0 & \cos(\theta_{y}) & i\sin(\theta_{y}) & 0 \\
   0 & i\sin(\theta_{y}) & \cos(\theta_{y}) & 0 \\
   -i\sin(\theta_{y}) & 0 & 0 & \cos(\theta_{y}) 
   \end{pmatrix} \\ \\
   e^{iJ_{z}t(\sigma_{1}^{z}\otimes\sigma_{2}^{z})/\hbar} =&  
   \begin{pmatrix}
   e^{i\theta_{z}} & 0 & 0 & 0 \\
   0 & e^{-i\theta_{z}} & 0 & 0 \\
   0 & 0 & e^{-i\theta_{z}} & 0 \\
   0 & 0 & 0 & e^{i\theta_{z}}
   \end{pmatrix} 
\end{aligned}
\end{equation}
where, $\theta_{\alpha} = tJ_{\alpha}/\hbar$. For $N=3$, there are terms that do not commute. For instance, if $p_{12}$ represents the Heisenberg interaction (Eq. \eqref{eq:2spin}) between spins 1 and 2 and $p_{23}$ represents interaction between spins 2 and 3, then $p_{12}$ does not commute with $p_{23}$. 
As a result, one cannot decompose the time evolution operator as a product of two-body evolution operators. The Trotter decomposition can be used to rewrite the time evolution operator in terms of two-body components as follows,
\begin{equation}
\begin{aligned}
    e^{-i\hat{H}t/\hbar} =& \big[\big(\prod_{\alpha}
    e^{i\theta_{\alpha}(\sigma_{1}^{\alpha}\otimes\sigma_{2}^{\alpha}\otimes \mathbb{1})/n}\big)\times\big(\prod_{\alpha}
    e^{i\theta_{\alpha}(\mathbb{1}\otimes\sigma_{2}^{\alpha}\otimes\sigma_{3}^{\alpha})/n}\big)\big]^{n} \\
    & + \mathcal{O}(t/n)
\end{aligned}
\end{equation}
where the error scales linearly with time step i.e. $t/n$, which can be a significant source of error. This can be mitigated by taking a smaller step size. However, this results in an overall increase in the computation cost. Therefore, there is a need to balance accuracy and computation cost.

\begin{figure}[h!]
    \centering
    \includegraphics[scale=0.4]{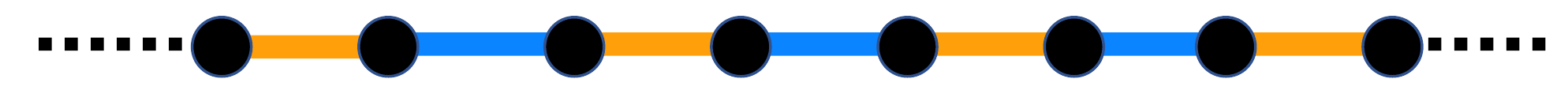}
    \caption{1D-spin chain showing two families (orange and blue) spin interaction which commute within family. }
    \label{fig:1Dchain}
\end{figure}

Extending the evolution operator to systems with $N>3$, one can observed that there are two major commuting families as shown in Fig.\ref{fig:1Dchain}. All elements in the orange family commute and all elements in blue family commute. Therefore the evolution operator can be written as a product of exponentials within the families without Trotter decomposition.

\subsection{Yang-Baxter Equation}

The Yang-Baxter equation (YBE) was introduced independently in theoretical physics by Yang~\cite{yang1967some} in the late 1960s, and by Baxter \cite{baxter1972partition} in statistical mechanics in the early 1970s. This relation has also received much attention in many areas of theoretical physics, classification of knots, scattering of subatomic particles, nuclear magnetic resonance, ultracold atoms, and more recently in quantum information science \cite{ge2016YBE,nayak2008non,kauffman2010topological,zhang2013integrable,vind2016experimental,batchelor2016yang}. 

The YBE connection to quantum computing originates from the pursuit of figuring out the relationship between topological entanglement, quantum entanglement, and quantum computational universality. In particular, how the global topological relationship in spaces (e.g. knotting and linking) corresponds to the entangled quantum states, and how the CNOT gate, for instance, can in turn be replaced by another unitary gate $R$ to maintain universality. It turns out these unitary $R$ gates that serve to maintain the universality of quantum computation and also as solutions for the condition of topological braiding are unitary solutions to the YBEs \cite{baxter2016exactly}
Briefly, the relation is a consistency or exchange condition that allows one to factorize the interactions of three bodies into a sequence of pairwise interactions under certain conditions. Formally, this can be written as 
\begin{equation}
    (\mathcal{R}\otimes\mathbb{1})(\mathbb{1}\otimes \mathcal{R})(\mathcal{R}\otimes\mathbb{1}) = (\mathbb{1}\otimes \mathcal{R})(\mathcal{R}\otimes\mathbb{1})(\mathbb{1}\otimes \mathcal{R})
\end{equation}
where the $\mathcal{R}$ operator is a linear mapping $\mathcal{R}: V\otimes V \rightarrow V\otimes V$ defined as a two-fold tensor product generalizing the permutation of vector space $V$. This relation also yields a sufficiency condition for quantum integrability in one dimensional quantum systems, and provides a systematic approach to construct integrable models. 
Since a detailed discussion of this topic is beyond the scope of this paper, we refer the reader to more comprehensive works and reviews on the subject \cite{jimbo1989ybe,baxter2016exactly}.

\section{Circuit Representation of the Time Evolution Operator}\label{section:time_evolution}
Since the evolution operator is a unitary matrix, there exists a quantum circuit that can perform this operation efficiently on a quantum computer.
First, we will find the quantum circuit for two spins and later extend it to $N$ spins with nearest neighbour interactions in one dimension. Each spin can be mapped to a qubit and the evolution of the spin system can be mapped to a quantum circuit. 
Using Eq. (\ref{eq:2spin}) and (\ref{eq:decomposition})
{\small
\begin{equation}\label{eq:time_xyz}
\begin{aligned}
    \prod_{\alpha=x,y,z}
    &e^{iJ_{\alpha}t(\sigma_{1}^{\alpha}\otimes\sigma_{2}^{\alpha})/\hbar} =\\
    &\begin{pmatrix}
   e^{i\theta_{z}}\cos(\gamma) & 0 & 0 & ie^{i\theta_{z}}\sin(\gamma) \\
   0 & e^{-i\theta_{z}}\cos(\delta) & ie^{-i\theta_{z}}\sin(\delta) & 0 \\
   0 & ie^{-i\theta_{z}}\sin(\delta) & e^{-i\theta_{z}}\cos(\delta) & 0 \\
   ie^{i\theta_{z}}\sin(\gamma) & 0 & 0 & e^{i\theta_{z}}\cos(\gamma) 
   \end{pmatrix} 
\end{aligned}
\end{equation}
}where, $\gamma = \theta_{x}-\theta_{y}$ and $\delta = \theta_{x}+\theta_{y}$. Optimal circuit for the above matrix is 
{\small
\begin{equation}\label{eq:qc_2spin}
\begin{aligned}
    \prod_{\alpha}
    &e^{iJ_{\alpha}t(\sigma_{1}^{\alpha}\otimes\sigma_{2}^{\alpha})/\hbar} =\\
    &\Qcircuit @C=0.5em @R=.7em @!R{
    &  \ctrl{1} & \gate{R_x(2 \theta_x)} & \gate{H} &  \ctrl{1} & \gate{S} & \gate{H}
    & \ctrl{1} & \gate{R_x(-\pi/2)} & \qw\\ 
    & \targ    & \gate{R_z(-2\theta_z)} & \qw & \targ    &  \gate{R_z(-2\theta_{y})} & \qw  
    & \targ    & \gate{R_x(\pi/2)} & \qw
    } 
\end{aligned}
\end{equation}
}The evolution operator for any time step can be represented using the above circuit. In addition, it is also a constant depth circuit for each time step since the number of one- and two-qubit gates does not increase with time step. The quantum circuit for a spin chain with more than two spins in one dimension can be derived using Eq. (\ref{eq:qc_2spin}) and the Trotter decomposition. 

\begin{figure}[h!]
    \centering
    \includegraphics[scale=0.3]{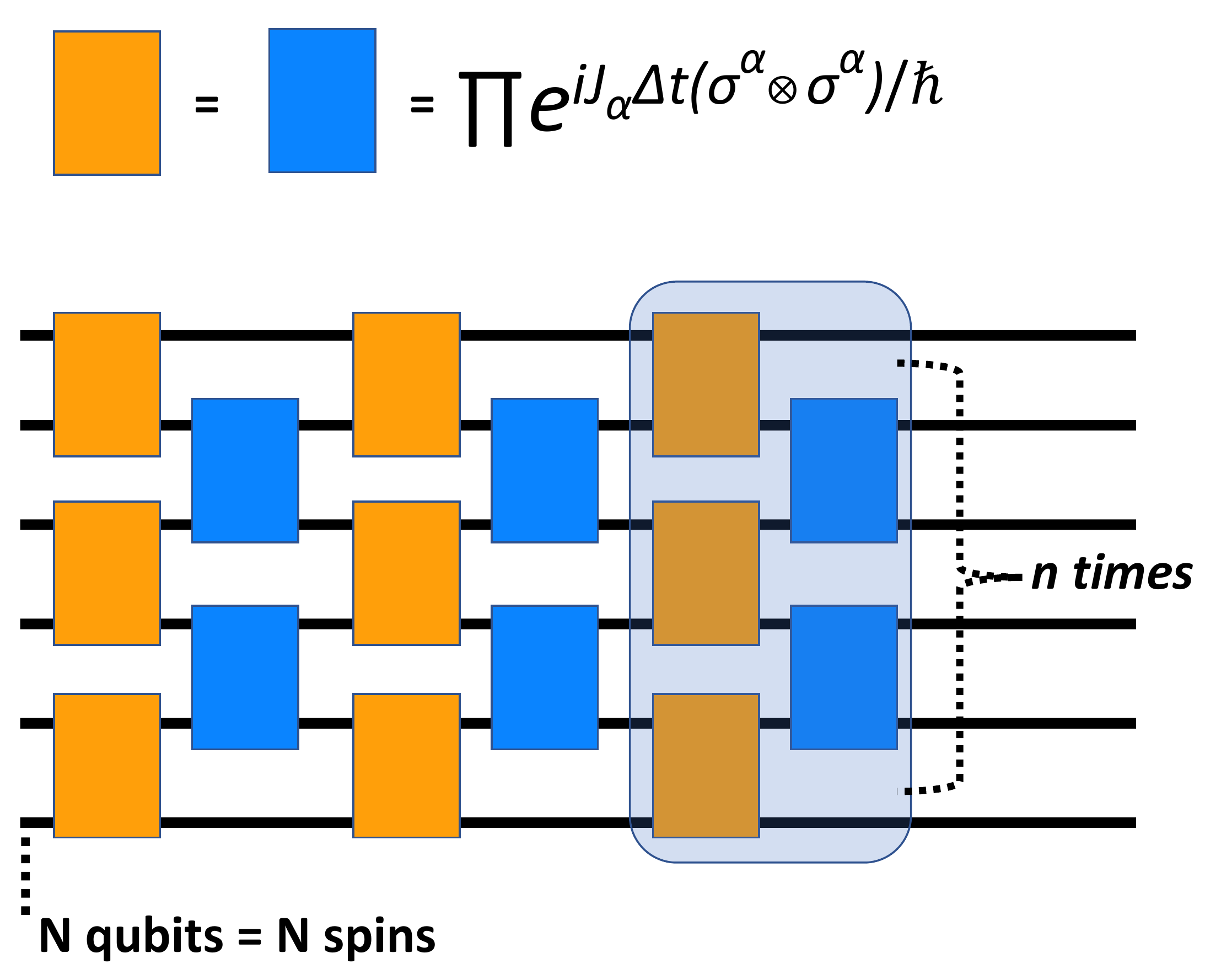}
    \caption{Quantum circuit for time evolution of $N$-spins, composed of $n$ alternative layers using the Trotter approximation.}
    \label{fig:1D_XYZ_QC}
\end{figure}

There exists two commuting families of operators as shown in Fig. \ref{fig:1D_XYZ_QC} as orange and blue two qubit gates, respectively. 
The accuracy of the simulation for a given time depends on the Trotter step ($t/n$). As a consequence of Trotterization, the quantum circuit for time evolution grows linearly with time step. Fig. \ref{fig:1D_XYZ_QC} shows the quantum circuit for a given time $t$ using $n$ Trotter steps. Each Trotter step is composed of a  bilayer of two-qubit gates. The first layer acts on the first two qubits, followed by the third and then the  fourth qubits and so on. Orange rectangles in Fig. \ref{fig:1D_XYZ_QC} represent the first layer. The second layer of two qubit gates starts from second qubit and acts on the next two qubits. Blue rectangles in Fig. \ref{fig:1D_XYZ_QC} represent the second layer. Both orange and blue rectangles combine to form an \emph{alternative layer}, covering all possible nearest neighbour interactions. 

\section{Circuit Compression Using the Yang-Baxter Equation}
In section \ref{section:time_evolution}, we showed the generalized circuit for time evolution of $N$ spins in one dimension. In this section we will utilize the YBE to simplify the generalized quantum circuit for arbitrary time step. First, we will show the existence of a unique reflection symmetry for a quantum circuit composed of alternative layers. Next, we will show the merge identity for two-qubit gates. We will also show how reflection symmetry combined with the merge identity allows for the compression of a quantum circuit of any length to a finite depth. 

\subsection{Reflection Symmetry and Merge Identity}
\begin{figure}[h!]
        \centering
        \includegraphics[scale=0.30]{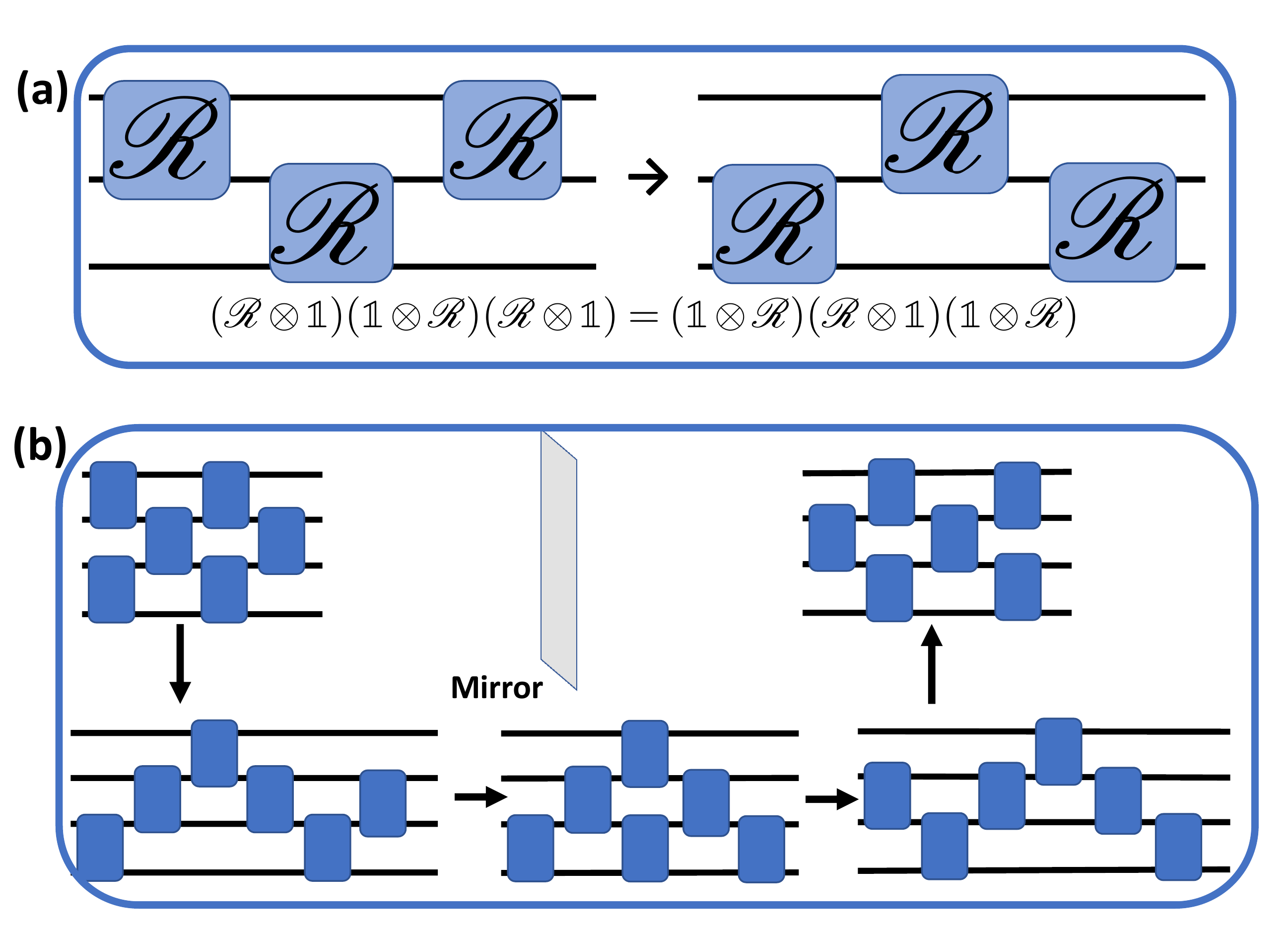}
        \caption{(a) Quantum circuit representation of the YBE for three qubits. (b) Reflection symmetry is achieved by using the YBE three times on four qubits.}
        \label{fig:reflection}
\end{figure}
The evolution operator for the Heisenberg Hamiltonian on two qubits is given by Eq. ( \ref{eq:time_xyz}). When there are two evolution operators with different parameters on the same two qubits, they can be merged and represented via a single operator as shown below
{\small
\begin{equation}\label{eq:merging}
    \mathcal{R}^{ij}(\theta_{x}^{1},\theta_{y}^{1},\theta_{z}^{1}).
    \mathcal{R}^{ij}(\theta_{x}^{2},\theta_{y}^{2},\theta_{z}^{2}) = 
    \mathcal{R}^{ij}(\theta_{x}^{1}+\theta_{x}^{2},\theta_{y}^{1}+\theta_{y}^{2},\theta_{z}^{1}+\theta_{z}^{2})
\end{equation}
}In rest of this article we shall call 
Eq. (\ref{eq:merging}) as the merge identity. Diagrammatic representation of the YBE is shown in the top panel of Fig. \ref{fig:reflection}, where $\mathcal{R}^{ij}$ represents the operator acting on $i,j$ qubits. By using this symmetry repeatedly one can prove the existence of reflection symmetry in $n$ qubits composed of $n/2$ alternative layers~ \cite{gulaniareflection}. The bottom panel of Fig. \ref{fig:reflection} shows how reflection symmetry is achieved for four qubits using the YBE three times. 

\begin{figure}[h!]
        \centering
        \includegraphics[scale=0.3]{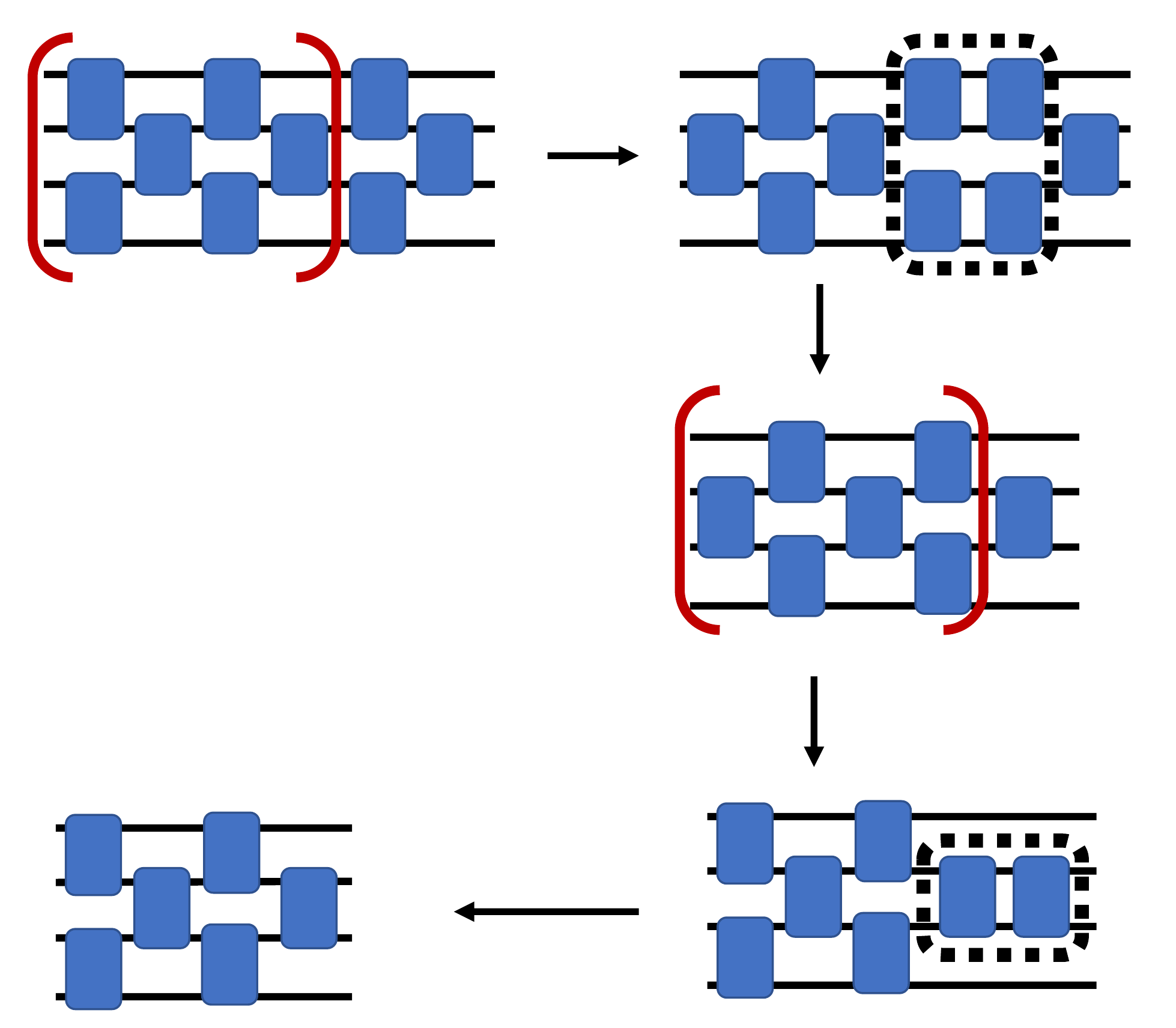}
        \caption{Compression scheme for 4 qubits. Reflection symmetry exists with two layers of alternative gates. Addition of a third layer can be absorbed into the two layers by recursive usage of reflection symmetry (red bracket) via the YBE and merge identity (black dotted box).}
        \label{fig:example}
\end{figure}

Reflection symmetry combined with the merge identity allows for the compression of $N$ alternative layers of gates to $N/2$ alternative layers for $N$ qubits. 
Fig. \ref{fig:example} shows the use of reflection symmetry combined with the merge identity for four qubits. A third alternative layer can be merged into the previous two layers. Therefore, any number of  alternative layers can be compressed into two alternative layers. 

\subsection{Algebraic Condition for Reflection Symmetry}

In the previous section, we have shown that reflection symmetry is a sufficient condition for performing circuit compression. An interesting follow-up question would be ``How would one know if reflection symmetry can be applied to a quantum dynamics simulation of a given Hamiltonian?'' To answer this question, one needs to show if there exist algebraic relations of phases and rotations before and after the reflection. However, given a general time propagator such as (\ref{eq:time_xyz}), an exhaustive search and rigorous proof appears challenging. Due to the lack of rigorous algebraic relations, YBE-like relations can only be conjectured as shown in previous work (see for example Ref. \citenum{bassman2021constantdepth}), which makes the compression of the circuit a heuristic process. In this section, we show that some algebraic relations can be obtained rigorously for at least  a few special cases of (\ref{eq:time_xyz}). In particular, we propose the following theorem. \\

\noindent \textbf{Theorem I} Given the time evolution operator that takes the following form
\begin{align}
\mathcal{R}(\gamma,\delta) & =  
\begin{pmatrix}
   e^{i\delta}\cos(\gamma) & 0 & 0 & ie^{i\delta}\sin(\gamma) \\
   0 & e^{-i\delta}\cos{\gamma} & ie^{-i\delta}\sin{\gamma} & 0 \\
   0 & ie^{-i\delta}\sin{\gamma} & e^{-i\delta}\cos{\gamma} & 0 \\
   ie^{i\delta}\sin(\gamma) & 0 & 0 & e^{i\delta}\cos(\gamma)
   \end{pmatrix}, \label{propagator}
\end{align}
then the following YBE holds
\begin{align}
    &(\mathcal{R}(\gamma_1,\delta_1) \otimes \mathbb{1})
    (\mathbb{1} \otimes \mathcal{R}(\gamma_2,\delta_2))
    (\mathcal{R}(\gamma_3,\delta_3) \otimes \mathbb{1})  \notag \\
    &~~~~=(\mathbb{1} \otimes \mathcal{R}(\gamma_4,\delta_4))
    (\mathcal{R}(\gamma_5,\delta_5) \otimes \mathbb{1})
    (\mathbb{1} \otimes \mathcal{R}(\gamma_6,\delta_6)) \label{YBE2}
\end{align}
if and only if the following 16 relations between the $\gamma$'s and $\delta$'s are satisfied
\begin{align}
 s_{\gamma_2} c_{\gamma_1-\gamma_3} c_{\delta_1-\delta_3} s_{\delta_2} &= c_{\gamma_5} s_{\gamma_4+\gamma_6} s_{\delta_4+\delta_6} c_{\delta_5} \label{r1}\\
 c_{\gamma_2} c_{\gamma_1-\gamma_3} c_{\delta_1+\delta_3} s_{\delta_2} &= c_{\gamma_5} c_{\gamma_4+\gamma_6} s_{\delta_4+\delta_6} c_{\delta_5} \label{r2},\\
-s_{\gamma_2} c_{\gamma_1+\gamma_3} s_{\delta_1-\delta_3} c_{\delta_2} &= c_{\gamma_5} s_{\gamma_4-\gamma_6} c_{\delta_4+\delta_6} s_{\delta_5} \label{r3},\\
 c_{\gamma_2} c_{\gamma_1+\gamma_3} s_{\delta_1+\delta_3} c_{\delta_2} &= c_{\gamma_5} c_{\gamma_4-\gamma_6} c_{\delta_4+\delta_6} s_{\delta_5} \label{r4},\\
 s_{\gamma_2} c_{\gamma_1+\gamma_3} c_{\delta_1-\delta_3} c_{\delta_2} &= c_{\gamma_5} s_{\gamma_4+\gamma_6} c_{\delta_4+\delta_6} c_{\delta_5} \label{r5},\\
 c_{\gamma_2} c_{\gamma_1+\gamma_3} c_{\delta_1+\delta_3} c_{\delta_2} &= c_{\gamma_5} c_{\gamma_4+\gamma_6} c_{\delta_4+\delta_6} c_{\delta_5} \label{r6},\\
-s_{\gamma_2} c_{\gamma_1-\gamma_3} s_{\delta_1-\delta_3} s_{\delta_2} &= c_{\gamma_5} s_{\gamma_4-\gamma_6} s_{\delta_4+\delta_6} s_{\delta_5} \label{r7},\\
 c_{\gamma_2} c_{\gamma_1-\gamma_3} s_{\delta_1+\delta_3} s_{\delta_2} &= c_{\gamma_5} c_{\gamma_4-\gamma_6} s_{\delta_4+\delta_6} s_{\delta_5} \label{r8},\\
 s_{\gamma_2} s_{\gamma_1+\gamma_3} c_{\delta_1-\delta_3} c_{\delta_2} &= s_{\gamma_5} s_{\gamma_4+\gamma_6} c_{\delta_4-\delta_6} c_{\delta_5} \label{r9},\\
 c_{\gamma_2} s_{\gamma_1+\gamma_3} c_{\delta_1+\delta_3} c_{\delta_2} &= s_{\gamma_5} c_{\gamma_4+\gamma_6} c_{\delta_4-\delta_6} c_{\delta_5} \label{r10},\\
 s_{\gamma_2} s_{\gamma_1-\gamma_3} s_{\delta_1-\delta_3} s_{\delta_2} &= s_{\gamma_5} s_{\gamma_4-\gamma_6} s_{\delta_4-\delta_6} s_{\delta_5} \label{r11},\\
-c_{\gamma_2} s_{\gamma_1-\gamma_3} s_{\delta_1+\delta_3} s_{\delta_2} &= s_{\gamma_5} c_{\gamma_4-\gamma_6} s_{\delta_4-\delta_6} s_{\delta_5} \label{r12},\\
-s_{\gamma_2} s_{\gamma_1-\gamma_3} c_{\delta_1-\delta_3} s_{\delta_2} &= s_{\gamma_5} s_{\gamma_4+\gamma_6} s_{\delta_4-\delta_6} c_{\delta_5} \label{r13},\\
-c_{\gamma_2} s_{\gamma_1-\gamma_3} c_{\delta_1+\delta_3} s_{\delta_2} &= s_{\gamma_5} c_{\gamma_4+\gamma_6} s_{\delta_4-\delta_6} c_{\delta_5} \label{r14},\\
-s_{\gamma_2} s_{\gamma_1+\gamma_3} s_{\delta_1-\delta_3} c_{\delta_2} &= s_{\gamma_5} s_{\gamma_4-\gamma_6} c_{\delta_4-\delta_6} s_{\delta_5} \label{r15},\\
 c_{\gamma_2} s_{\gamma_1+\gamma_3} s_{\delta_1+\delta_3} c_{\delta_2} &= s_{\gamma_5} c_{\gamma_4-\gamma_6} c_{\delta_4-\delta_6} s_{\delta_5} \label{r16},
\end{align}
where $s_{p}$ and $c_{p}$ denote $\sin{(p/2)}$ and $\cos{(p/2)}$, respectively. \\

The proof of $\textbf{Theorem I}$ is straightforward if we expand both sides of Eq. (\ref{YBE2}) and perform a term-by-term comparison. Based on this theorem,
we show explicitly in Table. \ref{tab:YBE_heisenberg}, where we show YBE analysis for six special Heisenberg Hamiltonians where reflection is accomplished algebraically. 
Here, the Hamiltonian operator $\hat{H}$ takes at most two terms from the set $\{H_X, H_Y, H_Z\}$ where 
\begin{align}
    H_{X} &=- \sum_{j=1}^{n-1} J_{x} \sigma^{x}_{j}\sigma^{x}_{j+1}, \\
    H_{Y} &=- \sum_{j=1}^{n-1} J_{y} \sigma^{y}_{j}\sigma^{y}_{j+1}, \\
    H_{Z} &=- \sum_{j=1}^{n-1} J_{z} \sigma^{z}_{j}\sigma^{z}_{j+1}.
\end{align}


\begin{table*}[]
\begin{tabular}{clcc}
\hline \hline
\multirow{2}{*}{Hamiltonian}    & \multirow{2}{*}{Time Propagator}       & \multirow{2}{*}{Time Propagator Circuit}       & \multirow{2}{*}{Necessary Conditions for YBE}    \\ 
&&&\\\hline
&&&\\
\multirow{2}{*}{$H_X$} & \multirow{2}{*}{$e^{- i t H_{X}/\hbar} = \mathcal{R}(\gamma,\delta=0)$} & \multirow{2}{*}{
\Qcircuit @C=0.5em @R=.7em @!R{
        &  \ctrl{1} & \gate{R_x(-2 \gamma)} & \ctrl{1} & \qw
         \\ 
        & \targ    & \qw & \targ  & \qw 
        }}
& 
$s_{\gamma_2} c_{\gamma_1+\gamma_3} =  c_{\gamma_5} s_{\gamma_4+\gamma_6}$~,~$c_{\gamma_2} c_{\gamma_1+\gamma_3} = c_{\gamma_5} c_{\gamma_4+\gamma_6}$ \\
&&&$s_{\gamma_2} s_{\gamma_1+\gamma_3} = s_{\gamma_5} s_{\gamma_4+\gamma_6}$~,~$c_{\gamma_2} s_{\gamma_1+\gamma_3} = s_{\gamma_5} c_{\gamma_4+\gamma_6}$ \\
&&&\\
&&&\\
\multirow{2}{*}{$H_Y$} & \multirow{2}{*}{$e^{- i t H_{Y}/\hbar} = U_1\mathcal{R}(\gamma,\delta=0)U_1^\dagger$} & \multirow{2}{*}{\Qcircuit @C=0.5em @R=.7em @!R{
        & \gate{R_z(\pi/2)}  &  \ctrl{1} & \gate{R_x(-2 \gamma)} & \ctrl{1} & \gate{R_z(-\pi/2)}& \qw
         \\ 
        & \gate{R_z(\pi/2)} & \targ    & \qw & \targ  & \gate{R_z(-\pi/2)}& \qw 
        }}
& \multirow{2}{*}{\text{Same as above}} \\
&&&\\
&&&\\
&&&\\
\multirow{2}{*}{$H_Z$} & \multirow{2}{*}{$e^{- i t H_{Z}/\hbar} = \mathcal{R}(\gamma=0,\delta)$} & \multirow{2}{*}{\Qcircuit @C=0.5em @R=.7em @!R{
        &  \ctrl{1} & \qw & \ctrl{1} & \qw
         \\ 
        & \targ    & \gate{R_z(-2 \delta)} & \targ  & \qw 
        }}
&  $c_{\delta_1+\delta_3} s_{\delta_2} = s_{\delta_4+\delta_6} c_{\delta_5}$~,~$s_{\delta_1+\delta_3} c_{\delta_2} = c_{\delta_4+\delta_6} s_{\delta_5}$
\\
&&&$c_{\delta_1+\delta_3} c_{\delta_2} = c_{\delta_4+\delta_6} c_{\delta_5}$~,~$s_{\delta_1+\delta_3} s_{\delta_2} = s_{\delta_4+\delta_6} s_{\delta_5}$\\
&&&\\
&&&\\
\multirow{2}{*}{$H_X+H_Y$} & $e^{- i t (H_{X}+H_{Y})/\hbar} = U_2\mathcal{R}(\tau,\phi)U_2^\dagger$ & \multirow{2}{*}{\Qcircuit @C=0.5em @R=.7em @!R{
    & \gate{R_x(\pi/2)} &  \ctrl{1} & \gate{R_x(- 2\tau)} & \ctrl{1} &  \gate{R_x(-\pi/2)}& \qw
    \\ 
    & \gate{R_x(\pi/2)} & \targ    & \gate{R_z(- 2\phi)} & \targ  & \gate{R_x(-\pi/2)}& \qw
    } }
& \multirow{2}{*}{($\tau,\phi$)-relation is analogous to Eqs. (\ref{r1}-\ref{r16})} \\
&$\tau = \gamma + \delta, \phi=\gamma-\delta$&&\\
&&&\\
&&&\\
\multirow{2}{*}{$H_X+H_Z$} & \multirow{2}{*}{$e^{- i t (H_{X}+H_{Z})/\hbar} = \mathcal{R}(\gamma,\delta)$} & \multirow{2}{*}{\Qcircuit @C=0.5em @R=.7em @!R{
    &  \ctrl{1} & \gate{R_x(-2 \gamma)} & \ctrl{1} & \qw
    \\ 
    & \targ    & \gate{R_z(-2 \delta)} & \targ  & \qw
    }   }
& \multirow{2}{*}{Eqs. (\ref{r1}-\ref{r16})} \\
&&&\\
&&&\\
&&&\\
\multirow{2}{*}{$H_Y+H_Z$} & \multirow{2}{*}{$e^{- i t (H_{Y}+H_{Z})/\hbar} = U_1\mathcal{R}(\gamma,\delta)U_1^\dagger$} & \multirow{2}{*}{\Qcircuit @C=0.5em @R=.7em @!R{
    & \gate{R_z(\pi/2)} &  \ctrl{1} & \gate{R_x(-2 \gamma)} & \ctrl{1} &  \gate{R_z(-\pi/2)}& \qw
    \\ 
    & \gate{R_z(\pi/2)} & \targ    & \gate{R_z(-2 \delta} & \targ  & \gate{R_z(-\pi/2)}& \qw
    }   }
& \multirow{2}{*}{Eqs. (\ref{r1}-\ref{r16})} \\
&&&\\
&&&\\ 
&&&\\ \hline \hline
\end{tabular}
\caption{YBE analysis for six special Heisenberg Hamiltonians. $U_1 = R_z(\pi/2)\otimes R_z(\pi/2)$ and $U_2 = R_x(\pi/2)\otimes R_x(\pi/2)$.}\label{tab:YBE_heisenberg}
\end{table*}

In practice, the algebraic relations between rotations and phases can be further simplified since we only care about one solution (not all the solutions) that satisfies these algebraic relations. For example, for $\hat{H} = H_X$, an apparent solution is to let $\gamma_6 =0$ (such that $\gamma_4 = \gamma_2$) and $\gamma_5 = \gamma_1 + \gamma_3$. Also, for $\hat{H} = H_X + H_Z$, if we don't consider the edge cases (that leads to singularities in the sine and cosine functions), ($\gamma_i,\delta_i$) ($i=4,5,6$) can be obtained from the following simplified trigonometric relations
\begin{align}
    \tan([\gamma_{4}+\gamma_{6}]/2) &= \tan(\gamma_{2}/2)\frac{\cos([\delta_{1}-\delta_{3}]/2)}{\cos([\delta_{1}+\delta_{3}]/2)}, \\
    \tan([\gamma_{4}-\gamma_{6}]/2) &= -\tan(\gamma_{2}/2)\frac{\sin([\delta_{1}-\delta_{3}]/2)}{\sin([\delta_{1}+\delta_{3}]/2)},\\
    \tan([\delta_{4}+\delta_{6}]/2) &= \tan(\delta_{2}/2)\frac{\cos([\gamma{1}-\gamma{3}]/2)}{\cos([\gamma{1}+\gamma{3}]/2)},  \\
    \tan([\delta_{4}-\delta_{6}]/2) &= \tan(\delta_{2}/2)\frac{\sin([\gamma{1}-\gamma{3}]/2)}{\sin([\gamma{1}+\gamma{3}]/2)}, \\
    \tan(\gamma_{5}/2) &= \tan([\gamma_{1}+\gamma_{3}]/2) \frac{\cos([\delta_{4}+\delta_{6}]/2)}{\cos([\delta_{4}-\delta_{6}]/2)},\\
    \tan(\delta_{5}/2) &= -\tan([\delta_{1}+\delta_{3}]/2) \frac{\cos([\gamma_{4}+\gamma_{6}]/2)}{\cos([\gamma_{4}-\gamma_{6}]/2)}.
\end{align}
\section{Time Dynamics on a Quantum Device}

As a proof of concept and to highlight the impact of compressed circuits on a real noisy quantum device (IBM-Manila, average CNOT error $\sim$ $10^{-3}$, average readout error $\sim 10^{-2}$ and 8192 shots), we performed a time dynamics simulation of the XY Hamiltonian with three spins. We compute the time-dependent staggered magnetization, $m_s(t)$, which can be connected to the antiferromagnetism and ferrimagnetism in materials as follows,
\begin{equation}
    m_s(t) = \frac{1}{N} \sum_{i} (-1)^{i}\braket{\sigma_{z}(t)} 
\end{equation}
The initial state is the ground state (Ne\'el state) of XY Hamiltonian defined as $\Psi_{0} = \ket{\uparrow\downarrow\uparrow\downarrow...\uparrow\downarrow}$. The staggered magnetization of the Ne\'el state is one. We performed the time evolution for 2.5 units of time with a Trotter step size of 0.025 units.
\begin{figure}[h!]
    \centering
    \includegraphics[scale=0.65]{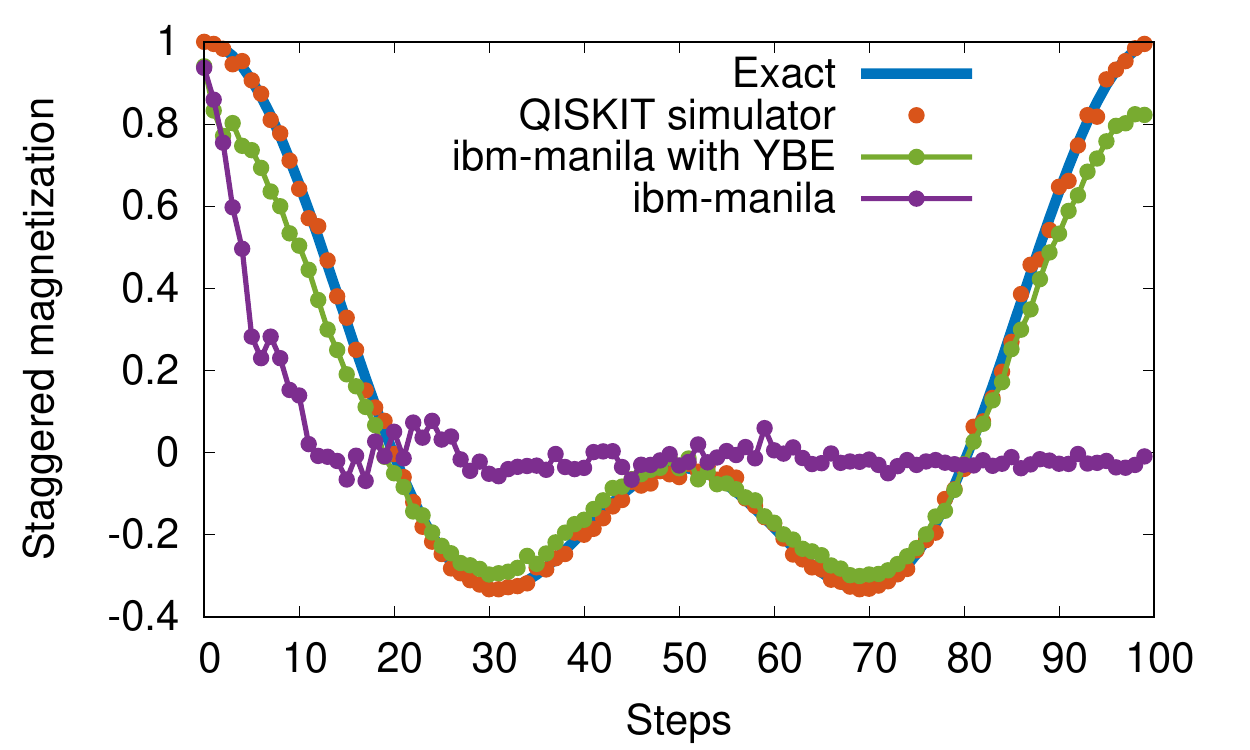}
    \caption{Comparison of time dynamics for 3 spin with the XY ($J_x = -0.8$ and $J_y = -0.2$) Heisenberg Hamiltonian on the IBM (Manila) device (8192 shots) with and without compression. Results from the Qiskit simulator serve as a baseline. }
    \label{fig:exp_4_qubits}
\end{figure}

Fig. \ref{fig:exp_4_qubits} shows the evolution of the staggered magnetization for three spins with parameters $J_x=-0.8$ and $J_y=-0.2$. We choose different parameters for $J_x$ and $J_y$ to make the system anisotropic.
The first component in the Fig. \ref{fig:exp_4_qubits} is the exact evolution of staggered magnetization for XY model, which serves as a reference. The second component is a simulation using the Qiskit simulator with no noise, which provides the estimation for running this evolution on a noise free quantum computer. Next component is the simulation result from the compressed circuit, which captures the dynamics for almost every time step. The compressed circuits are produced by repeated use of the YBE and merge identity. In contrast, results from a run on the same device with IBM-compiled circuits deviate quickly after the third time step. This comparison shows the impact of compressed circuits on a noisy quantum computer.

The compressed circuit simulation on a noisy quantum device shows an exceptional match with the exact evolution of staggered magnetization between steps 20-80. Also, the staggered magnetization difference at the zeroth and the last step is different. The amount of quantum error and its type are deeply rooted in both the observations. At the zeroth step, the error is small, which is a consequence of state preparation. However, at the last step, other errors from state evolution enter and make large deviations from exact staggered magnetization on top of state preparation error.  Depolarization noise converts a pure state to a maximally mixed state. Staggered magnetization for the maximally mixed state is zero. It is evident for uncompressed circuits, where staggered magnetization after the tenth step reaches zero and stays there for later steps. Therefore, for states which are have staggered magnetization zero, depolarization noise favors them. It is also the main contributor to the unparalleled overlap of simulation results from compressed circuits and exact evolution. Lastly, the compressed circuits will overlap more with exact evolution, irrespective of $J_x$ and $J_y$ due to small depth.

The results obtained from compressed circuits has controllable errors due to Trotter decomposition. The absolute error can be reduced with a smaller step size. However, a small step size increases the number of alternative layers, which can be compressed using our scheme to make it polynomial depth again. Therefore, our scheme allows for systematic convergence to the exact answer without increasing the circuit size. The only bottleneck is pre-processing of the circuit on a classical computer. Performing the YBE combined with the merge identity is computationally straightforward due to analytical expressions that we have derived. However, we note that multiple time usages of both can make the compression scheme computationally expensive for a large number of qubits. Computational complexity analysis for our compression scheme is currently being pursued and will be the subject of a future paper.

\section{Conclusions}
We have shown how the YBE can be utilized to compress and produce a shallow quantum circuit for efficient time dynamics simulations of 1-D lattice spin chains with nearest-neighbor interactions on real quantum computers. The depth of quantum circuits for each time step is independent of time, step size and only depends on the number of spins. Thus, the compressed circuit scales quadratically with system size, which allows for the simulations of time dynamics of very large 1-D spin chains. Moreover, we also derived the compressed circuit representations for different special cases of the Heisenberg Hamiltonian. To demonstrate the efficacy of the developed technique, we performed a time dynamics simulation of three spins on an IBM quantum computer and compared results from compressed and uncompressed quantum circuits, which confirmed the superiority of YBE formulation to perform dynamics for a large number of steps.

A promising application of this technique is to explore the compression of any circuit as part of the circuit compilation step. The general nature of the technique suggests that it could work for any circuit containing repeating gate motifs. It could be used, for example, to compress certain types of graph instances to solve combinatorial optimization problems using QAOA. In particular, certain $ZZ$ gate (a combination of $CNOT$, $Rz$, and $CNOT$ gates) motifs could be compressed.

This work can also pave the way to a demonstration of quantum advantage (a substantial computational advantage of quantum computing over classical) on NISQ devices. It has been proposed that quantum dynamics is a prime candidate. The reason why it has not been done yet is that NISQ devices are able to simulate circuits with very shallow depth. YBE technique solves this problem by making use of quantum computing to solve practical problems one step closer.



\section*{Acknowledgement}
This material is based upon work supported by the U.S. Department of Energy, Office of Science, National Quantum Information Science Research Centers. Y.A. acknowledges support from the U.S. Department of Energy, Office of Science, under contract DE-AC02-06CH11357 at Argonne National Laboratory. SG would like to thank QNEXT center for the opportunity and James Daniel Whitfield for insightful discussion. 


\bibliography{references}

\begin{thebibliography}{83}%
\makeatletter
\providecommand \@ifxundefined [1]{%
 \@ifx{#1\undefined}
}%
\providecommand \@ifnum [1]{%
 \ifnum #1\expandafter \@firstoftwo
 \else \expandafter \@secondoftwo
 \fi
}%
\providecommand \@ifx [1]{%
 \ifx #1\expandafter \@firstoftwo
 \else \expandafter \@secondoftwo
 \fi
}%
\providecommand \natexlab [1]{#1}%
\providecommand \enquote  [1]{``#1''}%
\providecommand \bibnamefont  [1]{#1}%
\providecommand \bibfnamefont [1]{#1}%
\providecommand \citenamefont [1]{#1}%
\providecommand \href@noop [0]{\@secondoftwo}%
\providecommand \href [0]{\begingroup \@sanitize@url \@href}%
\providecommand \@href[1]{\@@startlink{#1}\@@href}%
\providecommand \@@href[1]{\endgroup#1\@@endlink}%
\providecommand \@sanitize@url [0]{\catcode `\\12\catcode `\$12\catcode
  `\&12\catcode `\#12\catcode `\^12\catcode `\_12\catcode `\%12\relax}%
\providecommand \@@startlink[1]{}%
\providecommand \@@endlink[0]{}%
\providecommand \url  [0]{\begingroup\@sanitize@url \@url }%
\providecommand \@url [1]{\endgroup\@href {#1}{\urlprefix }}%
\providecommand \urlprefix  [0]{URL }%
\providecommand \Eprint [0]{\href }%
\providecommand \doibase [0]{https://doi.org/}%
\providecommand \selectlanguage [0]{\@gobble}%
\providecommand \bibinfo  [0]{\@secondoftwo}%
\providecommand \bibfield  [0]{\@secondoftwo}%
\providecommand \translation [1]{[#1]}%
\providecommand \BibitemOpen [0]{}%
\providecommand \bibitemStop [0]{}%
\providecommand \bibitemNoStop [0]{.\EOS\space}%
\providecommand \EOS [0]{\spacefactor3000\relax}%
\providecommand \BibitemShut  [1]{\csname bibitem#1\endcsname}%
\let\auto@bib@innerbib\@empty
\bibitem [{\citenamefont {Alexeev}\ \emph {et~al.}(2021)\citenamefont
  {Alexeev}, \citenamefont {Bacon}, \citenamefont {Brown}, \citenamefont
  {Calderbank}, \citenamefont {Carr}, \citenamefont {Chong}, \citenamefont
  {DeMarco}, \citenamefont {Englund}, \citenamefont {Farhi}, \citenamefont
  {Fefferman}, \citenamefont {Gorshkov}, \citenamefont {Houck}, \citenamefont
  {Kim}, \citenamefont {Kimmel}, \citenamefont {Lange}, \citenamefont {Lloyd},
  \citenamefont {Lukin}, \citenamefont {Maslov}, \citenamefont {Maunz},
  \citenamefont {Monroe}, \citenamefont {Preskill}, \citenamefont {Roetteler},
  \citenamefont {Savage},\ and\ \citenamefont {Thompson}}]{alexeev2021}%
  \BibitemOpen
  \bibfield  {author} {\bibinfo {author} {\bibfnamefont {Y.}~\bibnamefont
  {Alexeev}}, \bibinfo {author} {\bibfnamefont {D.}~\bibnamefont {Bacon}},
  \bibinfo {author} {\bibfnamefont {K.~R.}\ \bibnamefont {Brown}}, \bibinfo
  {author} {\bibfnamefont {R.}~\bibnamefont {Calderbank}}, \bibinfo {author}
  {\bibfnamefont {L.~D.}\ \bibnamefont {Carr}}, \bibinfo {author}
  {\bibfnamefont {F.~T.}\ \bibnamefont {Chong}}, \bibinfo {author}
  {\bibfnamefont {B.}~\bibnamefont {DeMarco}}, \bibinfo {author} {\bibfnamefont
  {D.}~\bibnamefont {Englund}}, \bibinfo {author} {\bibfnamefont
  {E.}~\bibnamefont {Farhi}}, \bibinfo {author} {\bibfnamefont
  {B.}~\bibnamefont {Fefferman}}, \bibinfo {author} {\bibfnamefont {A.~V.}\
  \bibnamefont {Gorshkov}}, \bibinfo {author} {\bibfnamefont {A.}~\bibnamefont
  {Houck}}, \bibinfo {author} {\bibfnamefont {J.}~\bibnamefont {Kim}}, \bibinfo
  {author} {\bibfnamefont {S.}~\bibnamefont {Kimmel}}, \bibinfo {author}
  {\bibfnamefont {M.}~\bibnamefont {Lange}}, \bibinfo {author} {\bibfnamefont
  {S.}~\bibnamefont {Lloyd}}, \bibinfo {author} {\bibfnamefont {M.~D.}\
  \bibnamefont {Lukin}}, \bibinfo {author} {\bibfnamefont {D.}~\bibnamefont
  {Maslov}}, \bibinfo {author} {\bibfnamefont {P.}~\bibnamefont {Maunz}},
  \bibinfo {author} {\bibfnamefont {C.}~\bibnamefont {Monroe}}, \bibinfo
  {author} {\bibfnamefont {J.}~\bibnamefont {Preskill}}, \bibinfo {author}
  {\bibfnamefont {M.}~\bibnamefont {Roetteler}}, \bibinfo {author}
  {\bibfnamefont {M.~J.}\ \bibnamefont {Savage}},\ and\ \bibinfo {author}
  {\bibfnamefont {J.}~\bibnamefont {Thompson}},\ }\bibfield  {title} {\bibinfo
  {title} {Quantum computer systems for scientific discovery},\ }\bibfield
  {journal} {\bibinfo  {journal} {{PRX} Quantum}\ }\textbf {\bibinfo {volume}
  {2}},\ \href {https://doi.org/10.1103/prxquantum.2.017001}
  {10.1103/prxquantum.2.017001} (\bibinfo {year} {2021})\BibitemShut {NoStop}%
\bibitem [{\citenamefont {Suzuki}(1976)}]{suzuki1976relationship}%
  \BibitemOpen
  \bibfield  {author} {\bibinfo {author} {\bibfnamefont {M.}~\bibnamefont
  {Suzuki}},\ }\bibfield  {title} {\bibinfo {title} {Relationship between
  d-dimensional quantal spin systems and (d+1)-dimensional ising systems:
  Equivalence, critical exponents and systematic approximants of the partition
  function and spin correlations},\ }\href@noop {} {\bibfield  {journal}
  {\bibinfo  {journal} {Progress of theoretical physics}\ }\textbf {\bibinfo
  {volume} {56}},\ \bibinfo {pages} {1454} (\bibinfo {year}
  {1976})}\BibitemShut {NoStop}%
\bibitem [{\citenamefont {Sachdev}(2011)}]{sachdev2011quantum}%
  \BibitemOpen
  \bibfield  {author} {\bibinfo {author} {\bibfnamefont {S.}~\bibnamefont
  {Sachdev}},\ }\href@noop {} {\emph {\bibinfo {title} {Quantum phase
  transitions}}}\ (\bibinfo  {publisher} {Cambridge university press},\
  \bibinfo {year} {2011})\BibitemShut {NoStop}%
\bibitem [{\citenamefont {Shankar}(2017)}]{shankar2017quantum}%
  \BibitemOpen
  \bibfield  {author} {\bibinfo {author} {\bibfnamefont {R.}~\bibnamefont
  {Shankar}},\ }\href@noop {} {\emph {\bibinfo {title} {Quantum Field Theory
  and Condensed Matter: An Introduction}}}\ (\bibinfo  {publisher} {Cambridge
  University Press},\ \bibinfo {year} {2017})\BibitemShut {NoStop}%
\bibitem [{\citenamefont {Sutherland}(2004)}]{sutherland2004beautiful}%
  \BibitemOpen
  \bibfield  {author} {\bibinfo {author} {\bibfnamefont {B.}~\bibnamefont
  {Sutherland}},\ }\href@noop {} {\emph {\bibinfo {title} {Beautiful models: 70
  years of exactly solved quantum many-body problems}}}\ (\bibinfo  {publisher}
  {World Scientific},\ \bibinfo {year} {2004})\BibitemShut {NoStop}%
\bibitem [{\citenamefont {Benioff}(1980)}]{benioff1980computer}%
  \BibitemOpen
  \bibfield  {author} {\bibinfo {author} {\bibfnamefont {P.}~\bibnamefont
  {Benioff}},\ }\bibfield  {title} {\bibinfo {title} {The computer as a
  physical system: A microscopic quantum mechanical hamiltonian model of
  computers as represented by turing machines},\ }\href@noop {} {\bibfield
  {journal} {\bibinfo  {journal} {Journal of Statistical Physics}\ }\textbf
  {\bibinfo {volume} {22}},\ \bibinfo {pages} {563} (\bibinfo {year}
  {1980})}\BibitemShut {NoStop}%
\bibitem [{\citenamefont {Feynman}(1982)}]{feynman1982simulating}%
  \BibitemOpen
  \bibfield  {author} {\bibinfo {author} {\bibfnamefont {R.~P.}\ \bibnamefont
  {Feynman}},\ }\bibfield  {title} {\bibinfo {title} {Simulating physics with
  computers},\ }\href@noop {} {\bibfield  {journal} {\bibinfo  {journal} {Int.
  J. Theor. Phys}\ }\textbf {\bibinfo {volume} {21}} (\bibinfo {year}
  {1982})}\BibitemShut {NoStop}%
\bibitem [{DOE(2021)}]{DOE-QIS}%
  \BibitemOpen
  \href@noop {} {\bibinfo {title} {{National QIS Research Centers, U.S. DOE
  Office of Science(SC)}}},\ \bibinfo {howpublished}
  {\url{https://science.osti.gov/Initiatives/QIS/QIS-Centers}} (\bibinfo {year}
  {2021}),\ \bibinfo {note} {[Online; accessed November, 17 2021]}\BibitemShut
  {NoStop}%
\bibitem [{\citenamefont {Preskill}(2018)}]{preskill2018quantum}%
  \BibitemOpen
  \bibfield  {author} {\bibinfo {author} {\bibfnamefont {J.}~\bibnamefont
  {Preskill}},\ }\bibfield  {title} {\bibinfo {title} {Quantum computing in the
  nisq era and beyond},\ }\href@noop {} {\bibfield  {journal} {\bibinfo
  {journal} {Quantum}\ }\textbf {\bibinfo {volume} {2}},\ \bibinfo {pages} {79}
  (\bibinfo {year} {2018})}\BibitemShut {NoStop}%
\bibitem [{\citenamefont {Jimbo}(1989)}]{jimbo1989ybe}%
  \BibitemOpen
  \bibfield  {author} {\bibinfo {author} {\bibfnamefont {M.}~\bibnamefont
  {Jimbo}},\ }\bibfield  {title} {\bibinfo {title} {Introduction to the
  yang-baxter equation},\ }\href@noop {} {\bibfield  {journal} {\bibinfo
  {journal} {International Journal of Modern Physics A}\ }\textbf {\bibinfo
  {volume} {4}},\ \bibinfo {pages} {3759} (\bibinfo {year} {1989})}\BibitemShut
  {NoStop}%
\bibitem [{\citenamefont {Bethe}(1931)}]{bethe1931theorie}%
  \BibitemOpen
  \bibfield  {author} {\bibinfo {author} {\bibfnamefont {H.}~\bibnamefont
  {Bethe}},\ }\bibfield  {title} {\bibinfo {title} {Zur theorie der metalle},\
  }\href {https://doi.org/10.1007/BF01341708} {\bibfield  {journal} {\bibinfo
  {journal} {Z. Physik}\ }\textbf {\bibinfo {volume} {71}},\ \bibinfo {pages}
  {205} (\bibinfo {year} {1931})}\BibitemShut {NoStop}%
\bibitem [{\citenamefont {Batchelor}(2007)}]{batchelor2007bethe}%
  \BibitemOpen
  \bibfield  {author} {\bibinfo {author} {\bibfnamefont {M.~T.}\ \bibnamefont
  {Batchelor}},\ }\bibfield  {title} {\bibinfo {title} {The bethe ansatz after
  75 years},\ }\href {https://doi.org/10.1063/1.2709557} {\bibfield  {journal}
  {\bibinfo  {journal} {Physics Today}\ }\textbf {\bibinfo {volume} {60}},\
  \bibinfo {pages} {36} (\bibinfo {year} {2007})},\ \Eprint
  {https://arxiv.org/abs/https://physicstoday.scitation.org/doi/pdf/10.1063/1.2709557}
  {https://physicstoday.scitation.org/doi/pdf/10.1063/1.2709557} \BibitemShut
  {NoStop}%
\bibitem [{\citenamefont {Yang}(1967)}]{yang1967some}%
  \BibitemOpen
  \bibfield  {author} {\bibinfo {author} {\bibfnamefont {C.-N.}\ \bibnamefont
  {Yang}},\ }\bibfield  {title} {\bibinfo {title} {Some exact results for the
  many-body problem in one dimension with repulsive delta-function
  interaction},\ }\href@noop {} {\bibfield  {journal} {\bibinfo  {journal}
  {Physical Review Letters}\ }\textbf {\bibinfo {volume} {19}},\ \bibinfo
  {pages} {1312} (\bibinfo {year} {1967})}\BibitemShut {NoStop}%
\bibitem [{\citenamefont {Baxter}(2016)}]{baxter2016exactly}%
  \BibitemOpen
  \bibfield  {author} {\bibinfo {author} {\bibfnamefont {R.~J.}\ \bibnamefont
  {Baxter}},\ }\href@noop {} {\emph {\bibinfo {title} {Exactly solved models in
  statistical mechanics}}}\ (\bibinfo  {publisher} {Elsevier},\ \bibinfo {year}
  {2016})\BibitemShut {NoStop}%
\bibitem [{\citenamefont {Perk}\ and\ \citenamefont
  {Au-Yang}(2006)}]{perk2006encyclopedia}%
  \BibitemOpen
  \bibfield  {author} {\bibinfo {author} {\bibfnamefont {J.}~\bibnamefont
  {Perk}}\ and\ \bibinfo {author} {\bibfnamefont {H.}~\bibnamefont {Au-Yang}},\
  }\href@noop {} {\bibinfo {title} {Encyclopedia of mathematical physics}}
  (\bibinfo {year} {2006})\BibitemShut {NoStop}%
\bibitem [{\citenamefont {Heisenberg}(1928)}]{heisenberg1928theorie}%
  \BibitemOpen
  \bibfield  {author} {\bibinfo {author} {\bibfnamefont {W.}~\bibnamefont
  {Heisenberg}},\ }\bibfield  {title} {\bibinfo {title} {Zur theorie des
  ferromagnetismus},\ }\href {https://doi.org/10.1007/BF01328601} {\bibfield
  {journal} {\bibinfo  {journal} {Z. Physik}\ }\textbf {\bibinfo {volume}
  {49}},\ \bibinfo {pages} {619} (\bibinfo {year} {1928})}\BibitemShut
  {NoStop}%
\bibitem [{\citenamefont {Takahashi}(1971)}]{takahashi1971one}%
  \BibitemOpen
  \bibfield  {author} {\bibinfo {author} {\bibfnamefont {M.}~\bibnamefont
  {Takahashi}},\ }\bibfield  {title} {\bibinfo {title} {{One-Dimensional
  Heisenberg Model at Finite Temperature}},\ }\href
  {https://doi.org/10.1143/PTP.46.401} {\bibfield  {journal} {\bibinfo
  {journal} {Progress of Theoretical Physics}\ }\textbf {\bibinfo {volume}
  {46}},\ \bibinfo {pages} {401} (\bibinfo {year} {1971})}\BibitemShut
  {NoStop}%
\bibitem [{\citenamefont {Karbach}\ \emph {et~al.}(1997)\citenamefont
  {Karbach}, \citenamefont {M\"{u}ller}, \citenamefont {Gould},\ and\
  \citenamefont {Tobochnik}}]{karbach1997introI}%
  \BibitemOpen
  \bibfield  {author} {\bibinfo {author} {\bibfnamefont {M.}~\bibnamefont
  {Karbach}}, \bibinfo {author} {\bibfnamefont {G.}~\bibnamefont {M\"{u}ller}},
  \bibinfo {author} {\bibfnamefont {H.}~\bibnamefont {Gould}},\ and\ \bibinfo
  {author} {\bibfnamefont {J.}~\bibnamefont {Tobochnik}},\ }\bibfield  {title}
  {\bibinfo {title} {Introduction to the bethe ansatz i},\ }\href
  {https://doi.org/10.1063/1.4822511} {\bibfield  {journal} {\bibinfo
  {journal} {Computers in Physics}\ }\textbf {\bibinfo {volume} {11}},\
  \bibinfo {pages} {36} (\bibinfo {year} {1997})}\BibitemShut {NoStop}%
\bibitem [{\citenamefont {Karbach}\ \emph {et~al.}(1998)\citenamefont
  {Karbach}, \citenamefont {Hu},\ and\ \citenamefont
  {M\"{u}ller}}]{karbach1998introII}%
  \BibitemOpen
  \bibfield  {author} {\bibinfo {author} {\bibfnamefont {M.}~\bibnamefont
  {Karbach}}, \bibinfo {author} {\bibfnamefont {K.}~\bibnamefont {Hu}},\ and\
  \bibinfo {author} {\bibfnamefont {G.}~\bibnamefont {M\"{u}ller}},\ }\bibfield
   {title} {\bibinfo {title} {Introduction to the bethe ansatz ii},\ }\href
  {https://doi.org/10.1063/1.168740} {\bibfield  {journal} {\bibinfo  {journal}
  {Computers in Physics}\ }\textbf {\bibinfo {volume} {12}},\ \bibinfo {pages}
  {565} (\bibinfo {year} {1998})}\BibitemShut {NoStop}%
\bibitem [{\citenamefont {Karbach}\ \emph {et~al.}(2000)\citenamefont
  {Karbach}, \citenamefont {Hu},\ and\ \citenamefont
  {M\"{u}ller}}]{karbach1998introIII}%
  \BibitemOpen
  \bibfield  {author} {\bibinfo {author} {\bibfnamefont {M.}~\bibnamefont
  {Karbach}}, \bibinfo {author} {\bibfnamefont {K.}~\bibnamefont {Hu}},\ and\
  \bibinfo {author} {\bibfnamefont {G.}~\bibnamefont {M\"{u}ller}},\
  }\href@noop {} {\bibinfo {title} {Introduction to the bethe ansatz iii}}
  (\bibinfo {year} {2000}),\ \Eprint {https://arxiv.org/abs/0008018}
  {arXiv:0008018 [cond-mat.stat-mech]} \BibitemShut {NoStop}%
\bibitem [{\citenamefont {Faddeev}(1996)}]{faddeev1996how}%
  \BibitemOpen
  \bibfield  {author} {\bibinfo {author} {\bibfnamefont {L.~D.}\ \bibnamefont
  {Faddeev}},\ }\href@noop {} {\bibinfo {title} {How algebraic bethe ansatz
  works for integrable model}} (\bibinfo {year} {1996}),\ \Eprint
  {https://arxiv.org/abs/9605187} {arXiv:9605187 [hep-th]} \BibitemShut
  {NoStop}%
\bibitem [{\citenamefont {Lieb}\ and\ \citenamefont
  {Liniger}(1963)}]{lieb1963exactI}%
  \BibitemOpen
  \bibfield  {author} {\bibinfo {author} {\bibfnamefont {E.~H.}\ \bibnamefont
  {Lieb}}\ and\ \bibinfo {author} {\bibfnamefont {W.}~\bibnamefont {Liniger}},\
  }\bibfield  {title} {\bibinfo {title} {Exact analysis of an interacting bose
  gas. i. the general solution and the ground state},\ }\href
  {https://doi.org/10.1103/PhysRev.130.1605} {\bibfield  {journal} {\bibinfo
  {journal} {Phys. Rev.}\ }\textbf {\bibinfo {volume} {130}},\ \bibinfo {pages}
  {1605} (\bibinfo {year} {1963})}\BibitemShut {NoStop}%
\bibitem [{\citenamefont {Lieb}(1963)}]{lieb1963exactII}%
  \BibitemOpen
  \bibfield  {author} {\bibinfo {author} {\bibfnamefont {E.~H.}\ \bibnamefont
  {Lieb}},\ }\bibfield  {title} {\bibinfo {title} {Exact analysis of an
  interacting bose gas. ii. the excitation spectrum},\ }\href
  {https://doi.org/10.1103/PhysRev.130.1616} {\bibfield  {journal} {\bibinfo
  {journal} {Phys. Rev.}\ }\textbf {\bibinfo {volume} {130}},\ \bibinfo {pages}
  {1616} (\bibinfo {year} {1963})}\BibitemShut {NoStop}%
\bibitem [{\citenamefont {Essler}\ \emph {et~al.}(2005)\citenamefont {Essler},
  \citenamefont {Frahm}, \citenamefont {G{\"o}hmann}, \citenamefont
  {Kl{\"u}mper},\ and\ \citenamefont {Korepin}}]{essler2005one}%
  \BibitemOpen
  \bibfield  {author} {\bibinfo {author} {\bibfnamefont {F.}~\bibnamefont
  {Essler}}, \bibinfo {author} {\bibfnamefont {H.}~\bibnamefont {Frahm}},
  \bibinfo {author} {\bibfnamefont {F.}~\bibnamefont {G{\"o}hmann}}, \bibinfo
  {author} {\bibfnamefont {A.}~\bibnamefont {Kl{\"u}mper}},\ and\ \bibinfo
  {author} {\bibfnamefont {V.}~\bibnamefont {Korepin}},\ }\href
  {https://books.google.com/books?id=wo0VPXtlK6oC} {\emph {\bibinfo {title}
  {The One-Dimensional Hubbard Model}}}\ (\bibinfo  {publisher} {Cambridge
  University Press},\ \bibinfo {year} {2005})\BibitemShut {NoStop}%
\bibitem [{\citenamefont {Calogero}(1969{\natexlab{a}})}]{calogero1969ground}%
  \BibitemOpen
  \bibfield  {author} {\bibinfo {author} {\bibfnamefont {F.}~\bibnamefont
  {Calogero}},\ }\bibfield  {title} {\bibinfo {title} {Ground state of a
  one‐dimensional n‐body system},\ }\href
  {https://doi.org/10.1063/1.1664821} {\bibfield  {journal} {\bibinfo
  {journal} {Journal of Mathematical Physics}\ }\textbf {\bibinfo {volume}
  {10}},\ \bibinfo {pages} {2197} (\bibinfo {year}
  {1969}{\natexlab{a}})}\BibitemShut {NoStop}%
\bibitem [{\citenamefont
  {Calogero}(1969{\natexlab{b}})}]{calogero1969solution}%
  \BibitemOpen
  \bibfield  {author} {\bibinfo {author} {\bibfnamefont {F.}~\bibnamefont
  {Calogero}},\ }\bibfield  {title} {\bibinfo {title} {Solution of a
  three‐body problem in one dimension},\ }\href
  {https://doi.org/10.1063/1.1664820} {\bibfield  {journal} {\bibinfo
  {journal} {Journal of Mathematical Physics}\ }\textbf {\bibinfo {volume}
  {10}},\ \bibinfo {pages} {2191} (\bibinfo {year}
  {1969}{\natexlab{b}})}\BibitemShut {NoStop}%
\bibitem [{\citenamefont {Calogero}(1971)}]{calogero1971solution}%
  \BibitemOpen
  \bibfield  {author} {\bibinfo {author} {\bibfnamefont {F.}~\bibnamefont
  {Calogero}},\ }\bibfield  {title} {\bibinfo {title} {Solution of the
  one‐dimensional n‐body problems with quadratic and/or inversely quadratic
  pair potentials},\ }\href {https://doi.org/10.1063/1.1665604} {\bibfield
  {journal} {\bibinfo  {journal} {Journal of Mathematical Physics}\ }\textbf
  {\bibinfo {volume} {12}},\ \bibinfo {pages} {419} (\bibinfo {year}
  {1971})}\BibitemShut {NoStop}%
\bibitem [{\citenamefont
  {Sutherland}(1971{\natexlab{a}})}]{sutherland1971quantumI}%
  \BibitemOpen
  \bibfield  {author} {\bibinfo {author} {\bibfnamefont {B.}~\bibnamefont
  {Sutherland}},\ }\bibfield  {title} {\bibinfo {title} {Quantum many‐body
  problem in one dimension: Ground state},\ }\href
  {https://doi.org/10.1063/1.1665584} {\bibfield  {journal} {\bibinfo
  {journal} {Journal of Mathematical Physics}\ }\textbf {\bibinfo {volume}
  {12}},\ \bibinfo {pages} {246} (\bibinfo {year}
  {1971}{\natexlab{a}})}\BibitemShut {NoStop}%
\bibitem [{\citenamefont
  {Sutherland}(1971{\natexlab{b}})}]{sutherland1971quantumII}%
  \BibitemOpen
  \bibfield  {author} {\bibinfo {author} {\bibfnamefont {B.}~\bibnamefont
  {Sutherland}},\ }\bibfield  {title} {\bibinfo {title} {Quantum many‐body
  problem in one dimension: Thermodynamics},\ }\href
  {https://doi.org/10.1063/1.1665585} {\bibfield  {journal} {\bibinfo
  {journal} {Journal of Mathematical Physics}\ }\textbf {\bibinfo {volume}
  {12}},\ \bibinfo {pages} {251} (\bibinfo {year}
  {1971}{\natexlab{b}})}\BibitemShut {NoStop}%
\bibitem [{\citenamefont {Johnson}\ \emph {et~al.}(1973)\citenamefont
  {Johnson}, \citenamefont {Krinsky},\ and\ \citenamefont
  {McCoy}}]{johnson1973vertical}%
  \BibitemOpen
  \bibfield  {author} {\bibinfo {author} {\bibfnamefont {J.~D.}\ \bibnamefont
  {Johnson}}, \bibinfo {author} {\bibfnamefont {S.}~\bibnamefont {Krinsky}},\
  and\ \bibinfo {author} {\bibfnamefont {B.~M.}\ \bibnamefont {McCoy}},\
  }\bibfield  {title} {\bibinfo {title} {Vertical-arrow correlation length in
  the eight-vertex model and the low-lying excitations of the
  $x\ensuremath{-}y\ensuremath{-}z$ hamiltonian},\ }\href
  {https://doi.org/10.1103/PhysRevA.8.2526} {\bibfield  {journal} {\bibinfo
  {journal} {Phys. Rev. A}\ }\textbf {\bibinfo {volume} {8}},\ \bibinfo {pages}
  {2526} (\bibinfo {year} {1973})}\BibitemShut {NoStop}%
\bibitem [{\citenamefont {Luther}(1976)}]{luther1976eigenvalue}%
  \BibitemOpen
  \bibfield  {author} {\bibinfo {author} {\bibfnamefont {A.}~\bibnamefont
  {Luther}},\ }\bibfield  {title} {\bibinfo {title} {Eigenvalue spectrum of
  interacting massive fermions in one dimension},\ }\href
  {https://doi.org/10.1103/PhysRevB.14.2153} {\bibfield  {journal} {\bibinfo
  {journal} {Phys. Rev. B}\ }\textbf {\bibinfo {volume} {14}},\ \bibinfo
  {pages} {2153} (\bibinfo {year} {1976})}\BibitemShut {NoStop}%
\bibitem [{\citenamefont {Orbach}(1958)}]{orbach1958linear}%
  \BibitemOpen
  \bibfield  {author} {\bibinfo {author} {\bibfnamefont {R.}~\bibnamefont
  {Orbach}},\ }\bibfield  {title} {\bibinfo {title} {Linear antiferromagnetic
  chain with anisotropic coupling},\ }\href
  {https://doi.org/10.1103/PhysRev.112.309} {\bibfield  {journal} {\bibinfo
  {journal} {Phys. Rev.}\ }\textbf {\bibinfo {volume} {112}},\ \bibinfo {pages}
  {309} (\bibinfo {year} {1958})}\BibitemShut {NoStop}%
\bibitem [{\citenamefont {Rigol}\ \emph {et~al.}(2007)\citenamefont {Rigol},
  \citenamefont {Dunjko}, \citenamefont {Yurovsky},\ and\ \citenamefont
  {Olshanii}}]{rigol2007relaxation}%
  \BibitemOpen
  \bibfield  {author} {\bibinfo {author} {\bibfnamefont {M.}~\bibnamefont
  {Rigol}}, \bibinfo {author} {\bibfnamefont {V.}~\bibnamefont {Dunjko}},
  \bibinfo {author} {\bibfnamefont {V.}~\bibnamefont {Yurovsky}},\ and\
  \bibinfo {author} {\bibfnamefont {M.}~\bibnamefont {Olshanii}},\ }\bibfield
  {title} {\bibinfo {title} {Relaxation in a completely integrable many-body
  quantum system: An ab initio study of the dynamics of the highly excited
  states of 1d lattice hard-core bosons},\ }\href
  {https://doi.org/10.1103/PhysRevLett.98.050405} {\bibfield  {journal}
  {\bibinfo  {journal} {Phys. Rev. Lett.}\ }\textbf {\bibinfo {volume} {98}},\
  \bibinfo {pages} {050405} (\bibinfo {year} {2007})}\BibitemShut {NoStop}%
\bibitem [{\citenamefont {Cassidy}\ \emph {et~al.}(2011)\citenamefont
  {Cassidy}, \citenamefont {Clark},\ and\ \citenamefont
  {Rigol}}]{cassidy2011generalized}%
  \BibitemOpen
  \bibfield  {author} {\bibinfo {author} {\bibfnamefont {A.~C.}\ \bibnamefont
  {Cassidy}}, \bibinfo {author} {\bibfnamefont {C.~W.}\ \bibnamefont {Clark}},\
  and\ \bibinfo {author} {\bibfnamefont {M.}~\bibnamefont {Rigol}},\ }\bibfield
   {title} {\bibinfo {title} {Generalized thermalization in an integrable
  lattice system},\ }\href {https://doi.org/10.1103/PhysRevLett.106.140405}
  {\bibfield  {journal} {\bibinfo  {journal} {Phys. Rev. Lett.}\ }\textbf
  {\bibinfo {volume} {106}},\ \bibinfo {pages} {140405} (\bibinfo {year}
  {2011})}\BibitemShut {NoStop}%
\bibitem [{\citenamefont {Caux}\ and\ \citenamefont
  {Essler}(2013)}]{caux2013time}%
  \BibitemOpen
  \bibfield  {author} {\bibinfo {author} {\bibfnamefont {J.-S.}\ \bibnamefont
  {Caux}}\ and\ \bibinfo {author} {\bibfnamefont {F.~H.~L.}\ \bibnamefont
  {Essler}},\ }\bibfield  {title} {\bibinfo {title} {Time evolution of local
  observables after quenching to an integrable model},\ }\href
  {https://doi.org/10.1103/PhysRevLett.110.257203} {\bibfield  {journal}
  {\bibinfo  {journal} {Phys. Rev. Lett.}\ }\textbf {\bibinfo {volume} {110}},\
  \bibinfo {pages} {257203} (\bibinfo {year} {2013})}\BibitemShut {NoStop}%
\bibitem [{\citenamefont {Fagotti}(2013)}]{fagotti2013dynamical}%
  \BibitemOpen
  \bibfield  {author} {\bibinfo {author} {\bibfnamefont {M.}~\bibnamefont
  {Fagotti}},\ }\href@noop {} {\bibinfo {title} {Dynamical phase transitions as
  properties of the stationary state: Analytic results after quantum quenches
  in the spin-1/2 xxz chain}} (\bibinfo {year} {2013}),\ \Eprint
  {https://arxiv.org/abs/1308.0277} {arXiv:1308.0277 [cond-mat.stat-mech]}
  \BibitemShut {NoStop}%
\bibitem [{\citenamefont {Pozsgay}(2013)}]{pozsgay2013generalized}%
  \BibitemOpen
  \bibfield  {author} {\bibinfo {author} {\bibfnamefont {B.}~\bibnamefont
  {Pozsgay}},\ }\bibfield  {title} {\bibinfo {title} {The generalized gibbs
  ensemble for heisenberg spin chains},\ }\href
  {https://doi.org/10.1088/1742-5468/2013/07/p07003} {\bibfield  {journal}
  {\bibinfo  {journal} {J. Stat. Mech.}\ }\textbf {\bibinfo {volume} {2013}},\
  \bibinfo {pages} {P07003} (\bibinfo {year} {2013})}\BibitemShut {NoStop}%
\bibitem [{\citenamefont {Sinitsyn}\ and\ \citenamefont
  {Li}(2016)}]{sinitsyn2016solvable}%
  \BibitemOpen
  \bibfield  {author} {\bibinfo {author} {\bibfnamefont {N.~A.}\ \bibnamefont
  {Sinitsyn}}\ and\ \bibinfo {author} {\bibfnamefont {F.}~\bibnamefont {Li}},\
  }\bibfield  {title} {\bibinfo {title} {Solvable multistate model of
  landau-zener transitions in cavity qed},\ }\href@noop {} {\bibfield
  {journal} {\bibinfo  {journal} {Physical Review A}\ }\textbf {\bibinfo
  {volume} {93}},\ \bibinfo {pages} {063859} (\bibinfo {year}
  {2016})}\BibitemShut {NoStop}%
\bibitem [{\citenamefont {Sinitsyn}\ \emph {et~al.}(2018)\citenamefont
  {Sinitsyn}, \citenamefont {Yuzbashyan}, \citenamefont {Chernyak},
  \citenamefont {Patra},\ and\ \citenamefont {Sun}}]{sinitsyn2018integrable}%
  \BibitemOpen
  \bibfield  {author} {\bibinfo {author} {\bibfnamefont {N.~A.}\ \bibnamefont
  {Sinitsyn}}, \bibinfo {author} {\bibfnamefont {E.~A.}\ \bibnamefont
  {Yuzbashyan}}, \bibinfo {author} {\bibfnamefont {V.~Y.}\ \bibnamefont
  {Chernyak}}, \bibinfo {author} {\bibfnamefont {A.}~\bibnamefont {Patra}},\
  and\ \bibinfo {author} {\bibfnamefont {C.}~\bibnamefont {Sun}},\ }\bibfield
  {title} {\bibinfo {title} {Integrable time-dependent quantum hamiltonians},\
  }\href@noop {} {\bibfield  {journal} {\bibinfo  {journal} {Physical review
  letters}\ }\textbf {\bibinfo {volume} {120}},\ \bibinfo {pages} {190402}
  (\bibinfo {year} {2018})}\BibitemShut {NoStop}%
\bibitem [{\citenamefont {Verstraete}\ \emph {et~al.}(2009)\citenamefont
  {Verstraete}, \citenamefont {Cirac},\ and\ \citenamefont
  {Latorre}}]{PhysRevA.79.032316}%
  \BibitemOpen
  \bibfield  {author} {\bibinfo {author} {\bibfnamefont {F.}~\bibnamefont
  {Verstraete}}, \bibinfo {author} {\bibfnamefont {J.~I.}\ \bibnamefont
  {Cirac}},\ and\ \bibinfo {author} {\bibfnamefont {J.~I.}\ \bibnamefont
  {Latorre}},\ }\bibfield  {title} {\bibinfo {title} {Quantum circuits for
  strongly correlated quantum systems},\ }\href
  {https://doi.org/10.1103/PhysRevA.79.032316} {\bibfield  {journal} {\bibinfo
  {journal} {Phys. Rev. A}\ }\textbf {\bibinfo {volume} {79}},\ \bibinfo
  {pages} {032316} (\bibinfo {year} {2009})}\BibitemShut {NoStop}%
\bibitem [{\citenamefont {Bassman}\ \emph {et~al.}(2021)\citenamefont
  {Bassman}, \citenamefont {Beeumen}, \citenamefont {Younis}, \citenamefont
  {Smith}, \citenamefont {Iancu},\ and\ \citenamefont
  {de~Jong}}]{bassman2021constantdepth}%
  \BibitemOpen
  \bibfield  {author} {\bibinfo {author} {\bibfnamefont {L.}~\bibnamefont
  {Bassman}}, \bibinfo {author} {\bibfnamefont {R.~V.}\ \bibnamefont
  {Beeumen}}, \bibinfo {author} {\bibfnamefont {E.}~\bibnamefont {Younis}},
  \bibinfo {author} {\bibfnamefont {E.}~\bibnamefont {Smith}}, \bibinfo
  {author} {\bibfnamefont {C.}~\bibnamefont {Iancu}},\ and\ \bibinfo {author}
  {\bibfnamefont {W.~A.}\ \bibnamefont {de~Jong}},\ }\href@noop {} {\bibinfo
  {title} {Constant-depth circuits for dynamic simulations of materials on
  quantum computers}} (\bibinfo {year} {2021}),\ \Eprint
  {https://arxiv.org/abs/2103.07429} {arXiv:2103.07429 [quant-ph]} \BibitemShut
  {NoStop}%
\bibitem [{\citenamefont {Fazekas}(1999)}]{fazekas1999lecture}%
  \BibitemOpen
  \bibfield  {author} {\bibinfo {author} {\bibfnamefont {P.}~\bibnamefont
  {Fazekas}},\ }\href@noop {} {\emph {\bibinfo {title} {Lecture notes on
  electron correlation and magnetism}}},\ Vol.~\bibinfo {volume} {5}\ (\bibinfo
   {publisher} {World scientific},\ \bibinfo {year} {1999})\BibitemShut
  {NoStop}%
\bibitem [{\citenamefont {Skomski}(2008)}]{skomski2008simple}%
  \BibitemOpen
  \bibfield  {author} {\bibinfo {author} {\bibfnamefont {R.}~\bibnamefont
  {Skomski}},\ }\href@noop {} {\emph {\bibinfo {title} {Simple models of
  magnetism}}}\ (\bibinfo  {publisher} {Oxford University Press},\ \bibinfo
  {year} {2008})\BibitemShut {NoStop}%
\bibitem [{\citenamefont {Pires}(2021)}]{pires2021theoretical}%
  \BibitemOpen
  \bibfield  {author} {\bibinfo {author} {\bibfnamefont {A.~S.~T.}\
  \bibnamefont {Pires}},\ }\href@noop {} {\emph {\bibinfo {title} {Theoretical
  Tools for Spin Models in Magnetic Systems}}}\ (\bibinfo  {publisher} {IOP
  Publishing},\ \bibinfo {year} {2021})\BibitemShut {NoStop}%
\bibitem [{\citenamefont {de~PR~Moreira}\ and\ \citenamefont
  {Illas}(2006)}]{illas2006unified}%
  \BibitemOpen
  \bibfield  {author} {\bibinfo {author} {\bibfnamefont {I.}~\bibnamefont
  {de~PR~Moreira}}\ and\ \bibinfo {author} {\bibfnamefont {F.}~\bibnamefont
  {Illas}},\ }\bibfield  {title} {\bibinfo {title} {A unified view of the
  theoretical description of magnetic coupling in molecular chemistry and solid
  state physics},\ }\href@noop {} {\bibfield  {journal} {\bibinfo  {journal}
  {Physical Chemistry Chemical Physics}\ }\textbf {\bibinfo {volume} {8}},\
  \bibinfo {pages} {1645} (\bibinfo {year} {2006})}\BibitemShut {NoStop}%
\bibitem [{\citenamefont {David}\ \emph {et~al.}(2017)\citenamefont {David},
  \citenamefont {Guih{\'e}ry},\ and\ \citenamefont
  {Ferré}}]{david2017physical}%
  \BibitemOpen
  \bibfield  {author} {\bibinfo {author} {\bibfnamefont {G.}~\bibnamefont
  {David}}, \bibinfo {author} {\bibfnamefont {N.}~\bibnamefont {Guih{\'e}ry}},\
  and\ \bibinfo {author} {\bibfnamefont {N.}~\bibnamefont {Ferré}},\
  }\bibfield  {title} {\bibinfo {title} {What are the physical contents of
  hubbard and heisenberg hamiltonian interactions extracted from broken
  symmetry dft calculations in magnetic compounds?},\ }\href@noop {} {\bibfield
   {journal} {\bibinfo  {journal} {Journal of chemical theory and computation}\
  }\textbf {\bibinfo {volume} {13}},\ \bibinfo {pages} {6253} (\bibinfo {year}
  {2017})}\BibitemShut {NoStop}%
\bibitem [{\citenamefont {Lieb}\ \emph {et~al.}(1961)\citenamefont {Lieb},
  \citenamefont {Schultz},\ and\ \citenamefont {Mattis}}]{lieb1961two}%
  \BibitemOpen
  \bibfield  {author} {\bibinfo {author} {\bibfnamefont {E.}~\bibnamefont
  {Lieb}}, \bibinfo {author} {\bibfnamefont {T.}~\bibnamefont {Schultz}},\ and\
  \bibinfo {author} {\bibfnamefont {D.}~\bibnamefont {Mattis}},\ }\bibfield
  {title} {\bibinfo {title} {Two soluble models of an antiferromagnetic
  chain},\ }\href
  {https://doi.org/https://doi.org/10.1016/0003-4916(61)90115-4} {\bibfield
  {journal} {\bibinfo  {journal} {Ann. Phys.}\ }\textbf {\bibinfo {volume}
  {16}},\ \bibinfo {pages} {407} (\bibinfo {year} {1961})}\BibitemShut
  {NoStop}%
\bibitem [{\citenamefont {Katsura}(1962)}]{katsura1962statistical}%
  \BibitemOpen
  \bibfield  {author} {\bibinfo {author} {\bibfnamefont {S.}~\bibnamefont
  {Katsura}},\ }\bibfield  {title} {\bibinfo {title} {Statistical mechanics of
  the anisotropic linear heisenberg model},\ }\href
  {https://doi.org/10.1103/PhysRev.127.1508} {\bibfield  {journal} {\bibinfo
  {journal} {Phys. Rev.}\ }\textbf {\bibinfo {volume} {127}},\ \bibinfo {pages}
  {1508} (\bibinfo {year} {1962})}\BibitemShut {NoStop}%
\bibitem [{\citenamefont {Katsura}(1963)}]{katsura1963statistical}%
  \BibitemOpen
  \bibfield  {author} {\bibinfo {author} {\bibfnamefont {S.}~\bibnamefont
  {Katsura}},\ }\bibfield  {title} {\bibinfo {title} {Statistical mechanics of
  the anisotropic linear heisenberg model},\ }\href
  {https://doi.org/10.1103/PhysRev.129.2835.4} {\bibfield  {journal} {\bibinfo
  {journal} {Phys. Rev.}\ }\textbf {\bibinfo {volume} {129}},\ \bibinfo {pages}
  {2835} (\bibinfo {year} {1963})}\BibitemShut {NoStop}%
\bibitem [{\citenamefont {Niemeijer}(1967)}]{niemeijer1967some}%
  \BibitemOpen
  \bibfield  {author} {\bibinfo {author} {\bibfnamefont {T.}~\bibnamefont
  {Niemeijer}},\ }\bibfield  {title} {\bibinfo {title} {Some exact calculations
  on a chain of spins 12},\ }\href
  {https://doi.org/https://doi.org/10.1016/0031-8914(67)90235-2} {\bibfield
  {journal} {\bibinfo  {journal} {Physica}\ }\textbf {\bibinfo {volume} {36}},\
  \bibinfo {pages} {377} (\bibinfo {year} {1967})}\BibitemShut {NoStop}%
\bibitem [{\citenamefont {Mehta}(1989)}]{mehta1989matrix}%
  \BibitemOpen
  \bibfield  {author} {\bibinfo {author} {\bibfnamefont {M.}~\bibnamefont
  {Mehta}},\ }\href {https://books.google.com/books?id=x9e0AAAACAAJ} {\emph
  {\bibinfo {title} {Matrix Theory: Selected Topics and Useful Results}}}\
  (\bibinfo  {publisher} {Hindustan Publishing Corporation},\ \bibinfo {year}
  {1989})\BibitemShut {NoStop}%
\bibitem [{\citenamefont {Shiroishi}\ \emph {et~al.}(2001)\citenamefont
  {Shiroishi}, \citenamefont {Takahashi},\ and\ \citenamefont
  {Nishiyama}}]{shiroishi2001emptiness}%
  \BibitemOpen
  \bibfield  {author} {\bibinfo {author} {\bibfnamefont {M.}~\bibnamefont
  {Shiroishi}}, \bibinfo {author} {\bibfnamefont {M.}~\bibnamefont
  {Takahashi}},\ and\ \bibinfo {author} {\bibfnamefont {Y.}~\bibnamefont
  {Nishiyama}},\ }\bibfield  {title} {\bibinfo {title} {Emptiness formation
  probability for the one-dimensional isotropic xy model},\ }\href
  {https://doi.org/10.1143/JPSJ.70.3535} {\bibfield  {journal} {\bibinfo
  {journal} {J. Phys. Soc. Jap.}\ }\textbf {\bibinfo {volume} {70}},\ \bibinfo
  {pages} {3535} (\bibinfo {year} {2001})}\BibitemShut {NoStop}%
\bibitem [{\citenamefont {Abanov}\ and\ \citenamefont
  {Franchini}(2003)}]{abanov2003emptiness}%
  \BibitemOpen
  \bibfield  {author} {\bibinfo {author} {\bibfnamefont {A.~G.}\ \bibnamefont
  {Abanov}}\ and\ \bibinfo {author} {\bibfnamefont {F.}~\bibnamefont
  {Franchini}},\ }\bibfield  {title} {\bibinfo {title} {Emptiness formation
  probability for the anisotropic xy spin chain in a magnetic field},\ }\href
  {https://doi.org/https://doi.org/10.1016/j.physleta.2003.07.009} {\bibfield
  {journal} {\bibinfo  {journal} {Phys. Lett. A}\ }\textbf {\bibinfo {volume}
  {316}},\ \bibinfo {pages} {342} (\bibinfo {year} {2003})}\BibitemShut
  {NoStop}%
\bibitem [{\citenamefont {Franchini}\ and\ \citenamefont
  {Abanov}(2005)}]{franchini2005asymptotics}%
  \BibitemOpen
  \bibfield  {author} {\bibinfo {author} {\bibfnamefont {F.}~\bibnamefont
  {Franchini}}\ and\ \bibinfo {author} {\bibfnamefont {A.~G.}\ \bibnamefont
  {Abanov}},\ }\bibfield  {title} {\bibinfo {title} {Asymptotics of toeplitz
  determinants and the emptiness formation probability for the {XY} spin
  chain},\ }\href {https://doi.org/10.1088/0305-4470/38/23/002} {\bibfield
  {journal} {\bibinfo  {journal} {J. Phys. A: Math. Theor.}\ }\textbf {\bibinfo
  {volume} {38}},\ \bibinfo {pages} {5069} (\bibinfo {year}
  {2005})}\BibitemShut {NoStop}%
\bibitem [{\citenamefont {Jin}\ and\ \citenamefont
  {Korepin}(2004)}]{jin2004quantum}%
  \BibitemOpen
  \bibfield  {author} {\bibinfo {author} {\bibfnamefont {B.-Q.}\ \bibnamefont
  {Jin}}\ and\ \bibinfo {author} {\bibfnamefont {V.~E.}\ \bibnamefont
  {Korepin}},\ }\bibfield  {title} {\bibinfo {title} {Quantum spin chain,
  toeplitz determinants and the fisher-hartwig conjecture},\ }\href
  {https://doi.org/10.1023/B:JOSS.0000037230.37166.42} {\bibfield  {journal}
  {\bibinfo  {journal} {J. Stat. Phys.}\ }\textbf {\bibinfo {volume} {116}},\
  \bibinfo {pages} {79} (\bibinfo {year} {2004})}\BibitemShut {NoStop}%
\bibitem [{\citenamefont {Its}\ \emph {et~al.}(2005)\citenamefont {Its},
  \citenamefont {Jin},\ and\ \citenamefont {Korepin}}]{its2005entanglement}%
  \BibitemOpen
  \bibfield  {author} {\bibinfo {author} {\bibfnamefont {A.~R.}\ \bibnamefont
  {Its}}, \bibinfo {author} {\bibfnamefont {B.-Q.}\ \bibnamefont {Jin}},\ and\
  \bibinfo {author} {\bibfnamefont {V.~E.}\ \bibnamefont {Korepin}},\
  }\bibfield  {title} {\bibinfo {title} {Entanglement in the xy spin chain},\
  }\href {https://doi.org/10.1088/0305-4470/38/13/011} {\bibfield  {journal}
  {\bibinfo  {journal} {J. Phys. A: Math. Theor.}\ }\textbf {\bibinfo {volume}
  {38}},\ \bibinfo {pages} {2975} (\bibinfo {year} {2005})}\BibitemShut
  {NoStop}%
\bibitem [{\citenamefont {A.~R.~Its}(2006)}]{its2006entropy}%
  \BibitemOpen
  \bibfield  {author} {\bibinfo {author} {\bibfnamefont {V.~E.~K.}\
  \bibnamefont {A.~R.~Its}, \bibfnamefont {B.-Q.~Jin}},\ }\href@noop {}
  {\bibinfo {title} {Entropy of xy spin chain and block toeplitz determinants}}
  (\bibinfo {year} {2006}),\ \Eprint {https://arxiv.org/abs/quant-ph/0606178}
  {arXiv:quant-ph/0606178 [quant-ph]} \BibitemShut {NoStop}%
\bibitem [{\citenamefont {Peschel}(2004)}]{peschel2004on}%
  \BibitemOpen
  \bibfield  {author} {\bibinfo {author} {\bibfnamefont {I.}~\bibnamefont
  {Peschel}},\ }\bibfield  {title} {\bibinfo {title} {On the entanglement
  entropy for an xy spin chain},\ }\href
  {https://doi.org/10.1088/1742-5468/2004/12/p12005} {\bibfield  {journal}
  {\bibinfo  {journal} {J. Stat. Mech.}\ }\textbf {\bibinfo {volume} {2004}},\
  \bibinfo {pages} {P12005} (\bibinfo {year} {2004})}\BibitemShut {NoStop}%
\bibitem [{\citenamefont {Franchini}\ \emph
  {et~al.}(2007{\natexlab{a}})\citenamefont {Franchini}, \citenamefont {Its},
  \citenamefont {Jin},\ and\ \citenamefont {Korepin}}]{franchini2007ellipses}%
  \BibitemOpen
  \bibfield  {author} {\bibinfo {author} {\bibfnamefont {F.}~\bibnamefont
  {Franchini}}, \bibinfo {author} {\bibfnamefont {A.~R.}\ \bibnamefont {Its}},
  \bibinfo {author} {\bibfnamefont {B.-Q.}\ \bibnamefont {Jin}},\ and\ \bibinfo
  {author} {\bibfnamefont {V.~E.}\ \bibnamefont {Korepin}},\ }\bibfield
  {title} {\bibinfo {title} {Ellipses of constant entropy in the xy spin
  chain},\ }\href {https://doi.org/10.1088/1751-8113/40/29/019} {\bibfield
  {journal} {\bibinfo  {journal} {J. Phys. A: Math. Theor.}\ }\textbf {\bibinfo
  {volume} {40}},\ \bibinfo {pages} {8467} (\bibinfo {year}
  {2007}{\natexlab{a}})}\BibitemShut {NoStop}%
\bibitem [{\citenamefont {Franchini}\ \emph
  {et~al.}(2007{\natexlab{b}})\citenamefont {Franchini}, \citenamefont {Its},\
  and\ \citenamefont {Korepin}}]{franchini2007renyi}%
  \BibitemOpen
  \bibfield  {author} {\bibinfo {author} {\bibfnamefont {F.}~\bibnamefont
  {Franchini}}, \bibinfo {author} {\bibfnamefont {A.~R.}\ \bibnamefont {Its}},\
  and\ \bibinfo {author} {\bibfnamefont {V.~E.}\ \bibnamefont {Korepin}},\
  }\bibfield  {title} {\bibinfo {title} {Renyi entropy of the {XY} spin
  chain},\ }\href {https://doi.org/10.1088/1751-8113/41/2/025302} {\bibfield
  {journal} {\bibinfo  {journal} {J. Phys. A: Math. Theor.}\ }\textbf {\bibinfo
  {volume} {41}},\ \bibinfo {pages} {025302} (\bibinfo {year}
  {2007}{\natexlab{b}})}\BibitemShut {NoStop}%
\bibitem [{\citenamefont {Silva}(2008)}]{silva2008statistics}%
  \BibitemOpen
  \bibfield  {author} {\bibinfo {author} {\bibfnamefont {A.}~\bibnamefont
  {Silva}},\ }\bibfield  {title} {\bibinfo {title} {Statistics of the work done
  on a quantum critical system by quenching a control parameter},\ }\href
  {https://doi.org/10.1103/PhysRevLett.101.120603} {\bibfield  {journal}
  {\bibinfo  {journal} {Phys. Rev. Lett.}\ }\textbf {\bibinfo {volume} {101}},\
  \bibinfo {pages} {120603} (\bibinfo {year} {2008})}\BibitemShut {NoStop}%
\bibitem [{\citenamefont {Calabrese}\ \emph {et~al.}(2011)\citenamefont
  {Calabrese}, \citenamefont {Essler},\ and\ \citenamefont
  {Fagotti}}]{calabrese2011quantum}%
  \BibitemOpen
  \bibfield  {author} {\bibinfo {author} {\bibfnamefont {P.}~\bibnamefont
  {Calabrese}}, \bibinfo {author} {\bibfnamefont {F.~H.~L.}\ \bibnamefont
  {Essler}},\ and\ \bibinfo {author} {\bibfnamefont {M.}~\bibnamefont
  {Fagotti}},\ }\bibfield  {title} {\bibinfo {title} {Quantum quench in the
  transverse-field ising chain},\ }\href
  {https://doi.org/10.1103/PhysRevLett.106.227203} {\bibfield  {journal}
  {\bibinfo  {journal} {Phys. Rev. Lett.}\ }\textbf {\bibinfo {volume} {106}},\
  \bibinfo {pages} {227203} (\bibinfo {year} {2011})}\BibitemShut {NoStop}%
\bibitem [{\citenamefont {Calabrese}\ \emph
  {et~al.}(2012{\natexlab{a}})\citenamefont {Calabrese}, \citenamefont
  {Essler},\ and\ \citenamefont {Fagotti}}]{calabrese2012quantumI}%
  \BibitemOpen
  \bibfield  {author} {\bibinfo {author} {\bibfnamefont {P.}~\bibnamefont
  {Calabrese}}, \bibinfo {author} {\bibfnamefont {F.~H.~L.}\ \bibnamefont
  {Essler}},\ and\ \bibinfo {author} {\bibfnamefont {M.}~\bibnamefont
  {Fagotti}},\ }\bibfield  {title} {\bibinfo {title} {Quantum quench in the
  transverse field ising chain: I. time evolution of order parameter
  correlators},\ }\href {https://doi.org/10.1088/1742-5468/2012/07/p07016}
  {\bibfield  {journal} {\bibinfo  {journal} {J. Stat. Mech.}\ }\textbf
  {\bibinfo {volume} {2012}},\ \bibinfo {pages} {P07016} (\bibinfo {year}
  {2012}{\natexlab{a}})}\BibitemShut {NoStop}%
\bibitem [{\citenamefont {Calabrese}\ \emph
  {et~al.}(2012{\natexlab{b}})\citenamefont {Calabrese}, \citenamefont
  {Essler},\ and\ \citenamefont {Fagotti}}]{calabrese2012quantumII}%
  \BibitemOpen
  \bibfield  {author} {\bibinfo {author} {\bibfnamefont {P.}~\bibnamefont
  {Calabrese}}, \bibinfo {author} {\bibfnamefont {F.~H.~L.}\ \bibnamefont
  {Essler}},\ and\ \bibinfo {author} {\bibfnamefont {M.}~\bibnamefont
  {Fagotti}},\ }\bibfield  {title} {\bibinfo {title} {Quantum quenches in the
  transverse field ising chain: Ii. stationary state properties},\ }\href
  {https://doi.org/10.1088/1742-5468/2012/07/p07022} {\bibfield  {journal}
  {\bibinfo  {journal} {J. Stat. Mech.}\ }\textbf {\bibinfo {volume} {2012}},\
  \bibinfo {pages} {P07022} (\bibinfo {year} {2012}{\natexlab{b}})}\BibitemShut
  {NoStop}%
\bibitem [{\citenamefont {H\"app\"ol\"a}\ \emph {et~al.}(2012)\citenamefont
  {H\"app\"ol\"a}, \citenamefont {Hal\'asz},\ and\ \citenamefont
  {Hamma}}]{happola2012universality}%
  \BibitemOpen
  \bibfield  {author} {\bibinfo {author} {\bibfnamefont {J.}~\bibnamefont
  {H\"app\"ol\"a}}, \bibinfo {author} {\bibfnamefont {G.~B.}\ \bibnamefont
  {Hal\'asz}},\ and\ \bibinfo {author} {\bibfnamefont {A.}~\bibnamefont
  {Hamma}},\ }\bibfield  {title} {\bibinfo {title} {Universality and robustness
  of revivals in the transverse field xy model},\ }\href
  {https://doi.org/10.1103/PhysRevA.85.032114} {\bibfield  {journal} {\bibinfo
  {journal} {Phys. Rev. A}\ }\textbf {\bibinfo {volume} {85}},\ \bibinfo
  {pages} {032114} (\bibinfo {year} {2012})}\BibitemShut {NoStop}%
\bibitem [{\citenamefont {Bucciantini}\ \emph {et~al.}(2014)\citenamefont
  {Bucciantini}, \citenamefont {Kormos},\ and\ \citenamefont
  {Calabrese}}]{bucciantini2014quantum}%
  \BibitemOpen
  \bibfield  {author} {\bibinfo {author} {\bibfnamefont {L.}~\bibnamefont
  {Bucciantini}}, \bibinfo {author} {\bibfnamefont {M.}~\bibnamefont
  {Kormos}},\ and\ \bibinfo {author} {\bibfnamefont {P.}~\bibnamefont
  {Calabrese}},\ }\bibfield  {title} {\bibinfo {title} {Quantum quenches from
  excited states in the ising chain},\ }\href
  {https://doi.org/10.1088/1751-8113/47/17/175002} {\bibfield  {journal}
  {\bibinfo  {journal} {J. Phys. A: Math. Theor.}\ }\textbf {\bibinfo {volume}
  {47}},\ \bibinfo {pages} {175002} (\bibinfo {year} {2014})}\BibitemShut
  {NoStop}%
\bibitem [{\citenamefont {Bayocboc}\ and\ \citenamefont
  {Paraan}(2015)}]{bayocboc2015exact}%
  \BibitemOpen
  \bibfield  {author} {\bibinfo {author} {\bibfnamefont {F.~A.}\ \bibnamefont
  {Bayocboc}}\ and\ \bibinfo {author} {\bibfnamefont {F.~N.~C.}\ \bibnamefont
  {Paraan}},\ }\bibfield  {title} {\bibinfo {title} {Exact work statistics of
  quantum quenches in the anisotropic xy model},\ }\href
  {https://doi.org/10.1103/PhysRevE.92.032142} {\bibfield  {journal} {\bibinfo
  {journal} {Phys. Rev. E}\ }\textbf {\bibinfo {volume} {92}},\ \bibinfo
  {pages} {032142} (\bibinfo {year} {2015})}\BibitemShut {NoStop}%
\bibitem [{\citenamefont {Barouch}\ \emph {et~al.}(1970)\citenamefont
  {Barouch}, \citenamefont {McCoy},\ and\ \citenamefont
  {Dresden}}]{barouch1970statI}%
  \BibitemOpen
  \bibfield  {author} {\bibinfo {author} {\bibfnamefont {E.}~\bibnamefont
  {Barouch}}, \bibinfo {author} {\bibfnamefont {B.~M.}\ \bibnamefont {McCoy}},\
  and\ \bibinfo {author} {\bibfnamefont {M.}~\bibnamefont {Dresden}},\
  }\bibfield  {title} {\bibinfo {title} {Statistical mechanics of the
  $\mathrm{XY}$ model. i},\ }\href {https://doi.org/10.1103/PhysRevA.2.1075}
  {\bibfield  {journal} {\bibinfo  {journal} {Phys. Rev. A}\ }\textbf {\bibinfo
  {volume} {2}},\ \bibinfo {pages} {1075} (\bibinfo {year} {1970})}\BibitemShut
  {NoStop}%
\bibitem [{\citenamefont {Barouch}\ and\ \citenamefont
  {McCoy}(1971{\natexlab{a}})}]{barouch1970statII}%
  \BibitemOpen
  \bibfield  {author} {\bibinfo {author} {\bibfnamefont {E.}~\bibnamefont
  {Barouch}}\ and\ \bibinfo {author} {\bibfnamefont {B.~M.}\ \bibnamefont
  {McCoy}},\ }\bibfield  {title} {\bibinfo {title} {Statistical mechanics of
  the $xy$ model. ii. spin-correlation functions},\ }\href
  {https://doi.org/10.1103/PhysRevA.3.786} {\bibfield  {journal} {\bibinfo
  {journal} {Phys. Rev. A}\ }\textbf {\bibinfo {volume} {3}},\ \bibinfo {pages}
  {786} (\bibinfo {year} {1971}{\natexlab{a}})}\BibitemShut {NoStop}%
\bibitem [{\citenamefont {Barouch}\ and\ \citenamefont
  {McCoy}(1971{\natexlab{b}})}]{barouch1970statIII}%
  \BibitemOpen
  \bibfield  {author} {\bibinfo {author} {\bibfnamefont {E.}~\bibnamefont
  {Barouch}}\ and\ \bibinfo {author} {\bibfnamefont {B.~M.}\ \bibnamefont
  {McCoy}},\ }\bibfield  {title} {\bibinfo {title} {Statistical mechanics of
  the $\mathrm{XY}$ model. iii},\ }\href
  {https://doi.org/10.1103/PhysRevA.3.2137} {\bibfield  {journal} {\bibinfo
  {journal} {Phys. Rev. A}\ }\textbf {\bibinfo {volume} {3}},\ \bibinfo {pages}
  {2137} (\bibinfo {year} {1971}{\natexlab{b}})}\BibitemShut {NoStop}%
\bibitem [{\citenamefont {McCoy}\ \emph {et~al.}(1971)\citenamefont {McCoy},
  \citenamefont {Barouch},\ and\ \citenamefont {Abraham}}]{mccoy1971statIV}%
  \BibitemOpen
  \bibfield  {author} {\bibinfo {author} {\bibfnamefont {B.~M.}\ \bibnamefont
  {McCoy}}, \bibinfo {author} {\bibfnamefont {E.}~\bibnamefont {Barouch}},\
  and\ \bibinfo {author} {\bibfnamefont {D.~B.}\ \bibnamefont {Abraham}},\
  }\bibfield  {title} {\bibinfo {title} {Statistical mechanics of the
  $\mathrm{XY}$ model. iv. time-dependent spin-correlation functions},\ }\href
  {https://doi.org/10.1103/PhysRevA.4.2331} {\bibfield  {journal} {\bibinfo
  {journal} {Phys. Rev. A}\ }\textbf {\bibinfo {volume} {4}},\ \bibinfo {pages}
  {2331} (\bibinfo {year} {1971})}\BibitemShut {NoStop}%
\bibitem [{\citenamefont {Franchini}(2017)}]{franchini2017introduction}%
  \BibitemOpen
  \bibfield  {author} {\bibinfo {author} {\bibfnamefont {F.}~\bibnamefont
  {Franchini}},\ }\href {https://books.google.com/books?id=UGclDwAAQBAJ} {\emph
  {\bibinfo {title} {An Introduction to Integrable Techniques for
  One-Dimensional Quantum Systems}}},\ Lecture Notes in Physics\ (\bibinfo
  {publisher} {Springer International Publishing},\ \bibinfo {year}
  {2017})\BibitemShut {NoStop}%
\bibitem [{\citenamefont {Mussardo}(2010)}]{mussardo2010statistical}%
  \BibitemOpen
  \bibfield  {author} {\bibinfo {author} {\bibfnamefont {G.}~\bibnamefont
  {Mussardo}},\ }\href {https://books.google.com/books?id=fakVDAAAQBAJ} {\emph
  {\bibinfo {title} {Statistical Field Theory: An Introduction to Exactly
  Solved Models in Statistical Physics}}},\ Oxford Graduate Texts\ (\bibinfo
  {publisher} {OUP Oxford},\ \bibinfo {year} {2010})\BibitemShut {NoStop}%
\bibitem [{\citenamefont {Kosloff}(1988)}]{kosloff1988time}%
  \BibitemOpen
  \bibfield  {author} {\bibinfo {author} {\bibfnamefont {R.}~\bibnamefont
  {Kosloff}},\ }\bibfield  {title} {\bibinfo {title} {Time-dependent
  quantum-mechanical methods for molecular dynamics},\ }\href@noop {}
  {\bibfield  {journal} {\bibinfo  {journal} {The Journal of Physical
  Chemistry}\ }\textbf {\bibinfo {volume} {92}},\ \bibinfo {pages} {2087}
  (\bibinfo {year} {1988})}\BibitemShut {NoStop}%
\bibitem [{\citenamefont {Tannor}(2007)}]{tannor2007introduction}%
  \BibitemOpen
  \bibfield  {author} {\bibinfo {author} {\bibfnamefont {D.~J.}\ \bibnamefont
  {Tannor}},\ }\href@noop {} {\emph {\bibinfo {title} {Introduction to quantum
  mechanics: a time-dependent perspective}}}\ (\bibinfo  {publisher}
  {University Science Books, Sausalito, CA},\ \bibinfo {year}
  {2007})\BibitemShut {NoStop}%
\bibitem [{\citenamefont {Baxter}(1972)}]{baxter1972partition}%
  \BibitemOpen
  \bibfield  {author} {\bibinfo {author} {\bibfnamefont {R.~J.}\ \bibnamefont
  {Baxter}},\ }\bibfield  {title} {\bibinfo {title} {Partition function of the
  eight-vertex lattice model},\ }\href@noop {} {\bibfield  {journal} {\bibinfo
  {journal} {Annals of Physics}\ }\textbf {\bibinfo {volume} {70}},\ \bibinfo
  {pages} {193} (\bibinfo {year} {1972})}\BibitemShut {NoStop}%
\bibitem [{\citenamefont {Ge}\ \emph {et~al.}(2016)\citenamefont {Ge},
  \citenamefont {Xue}, \citenamefont {Zhang},\ and\ \citenamefont
  {Zhao}}]{ge2016YBE}%
  \BibitemOpen
  \bibfield  {author} {\bibinfo {author} {\bibfnamefont {M.-L.}\ \bibnamefont
  {Ge}}, \bibinfo {author} {\bibfnamefont {K.}~\bibnamefont {Xue}}, \bibinfo
  {author} {\bibfnamefont {R.-Y.}\ \bibnamefont {Zhang}},\ and\ \bibinfo
  {author} {\bibfnamefont {Q.}~\bibnamefont {Zhao}},\ }\bibfield  {title}
  {\bibinfo {title} {Yang–baxter equations and quantum entanglements},\
  }\href {https://doi.org/10.1007/s11128-014-0765-3} {\bibfield  {journal}
  {\bibinfo  {journal} {Quantum Inf. Process.}\ }\textbf {\bibinfo {volume}
  {15}},\ \bibinfo {pages} {5211} (\bibinfo {year} {2016})}\BibitemShut
  {NoStop}%
\bibitem [{\citenamefont {Nayak}\ \emph {et~al.}(2008)\citenamefont {Nayak},
  \citenamefont {Simon}, \citenamefont {Stern}, \citenamefont {Freedman},\ and\
  \citenamefont {Das~Sarma}}]{nayak2008non}%
  \BibitemOpen
  \bibfield  {author} {\bibinfo {author} {\bibfnamefont {C.}~\bibnamefont
  {Nayak}}, \bibinfo {author} {\bibfnamefont {S.~H.}\ \bibnamefont {Simon}},
  \bibinfo {author} {\bibfnamefont {A.}~\bibnamefont {Stern}}, \bibinfo
  {author} {\bibfnamefont {M.}~\bibnamefont {Freedman}},\ and\ \bibinfo
  {author} {\bibfnamefont {S.}~\bibnamefont {Das~Sarma}},\ }\bibfield  {title}
  {\bibinfo {title} {Non-abelian anyons and topological quantum computation},\
  }\href {https://doi.org/10.1103/RevModPhys.80.1083} {\bibfield  {journal}
  {\bibinfo  {journal} {Rev. Mod. Phys.}\ }\textbf {\bibinfo {volume} {80}},\
  \bibinfo {pages} {1083} (\bibinfo {year} {2008})}\BibitemShut {NoStop}%
\bibitem [{\citenamefont {Kauffman}\ and\ \citenamefont
  {Lomonaco}(2010)}]{kauffman2010topological}%
  \BibitemOpen
  \bibfield  {author} {\bibinfo {author} {\bibfnamefont {L.~H.}\ \bibnamefont
  {Kauffman}}\ and\ \bibinfo {author} {\bibfnamefont {S.~J.~J.}\ \bibnamefont
  {Lomonaco}},\ }\bibfield  {title} {\bibinfo {title} {Topological quantum
  information theory},\ }in\ \href@noop {} {\emph {\bibinfo {booktitle}
  {Proceedings of Symposia in Applied Mathematics}}},\ \bibinfo {series and
  number} {Vol. 68},\ \bibinfo {editor} {edited by\ \bibinfo {editor}
  {\bibfnamefont {S.~J.}\ \bibnamefont {Lomonaco}}}\ (\bibinfo  {publisher}
  {AMS},\ \bibinfo {address} {Washington DC},\ \bibinfo {year}
  {2010})\BibitemShut {NoStop}%
\bibitem [{\citenamefont {Zhang}(2013)}]{zhang2013integrable}%
  \BibitemOpen
  \bibfield  {author} {\bibinfo {author} {\bibfnamefont {Y.}~\bibnamefont
  {Zhang}},\ }\bibfield  {title} {\bibinfo {title} {Integrable quantum
  computation},\ }\href {https://doi.org/10.1007/s11128-012-0409-4} {\bibfield
  {journal} {\bibinfo  {journal} {Quantum Inf. Process.}\ }\textbf {\bibinfo
  {volume} {12}},\ \bibinfo {pages} {631} (\bibinfo {year} {2013})}\BibitemShut
  {NoStop}%
\bibitem [{\citenamefont {Vind}\ \emph {et~al.}(2016)\citenamefont {Vind},
  \citenamefont {Foerster}, \citenamefont {Oliveira}, \citenamefont {Sarthour},
  \citenamefont {Soares-Pinto}, \citenamefont {Souza},\ and\ \citenamefont
  {Roditi}}]{vind2016experimental}%
  \BibitemOpen
  \bibfield  {author} {\bibinfo {author} {\bibfnamefont {F.~A.}\ \bibnamefont
  {Vind}}, \bibinfo {author} {\bibfnamefont {A.}~\bibnamefont {Foerster}},
  \bibinfo {author} {\bibfnamefont {I.~S.}\ \bibnamefont {Oliveira}}, \bibinfo
  {author} {\bibfnamefont {R.~S.}\ \bibnamefont {Sarthour}}, \bibinfo {author}
  {\bibfnamefont {D.~d.~O.}\ \bibnamefont {Soares-Pinto}}, \bibinfo {author}
  {\bibfnamefont {A.~M.~d.}\ \bibnamefont {Souza}},\ and\ \bibinfo {author}
  {\bibfnamefont {I.}~\bibnamefont {Roditi}},\ }\bibfield  {title} {\bibinfo
  {title} {Experimental realization of the yang-baxter equation via nmr
  interferometry},\ }\href@noop {} {\bibfield  {journal} {\bibinfo  {journal}
  {Scientific reports}\ }\textbf {\bibinfo {volume} {6}},\ \bibinfo {pages} {1}
  (\bibinfo {year} {2016})}\BibitemShut {NoStop}%
\bibitem [{\citenamefont {Batchelor}\ and\ \citenamefont
  {Foerster}(2016)}]{batchelor2016yang}%
  \BibitemOpen
  \bibfield  {author} {\bibinfo {author} {\bibfnamefont {M.~T.}\ \bibnamefont
  {Batchelor}}\ and\ \bibinfo {author} {\bibfnamefont {A.}~\bibnamefont
  {Foerster}},\ }\bibfield  {title} {\bibinfo {title} {Yang-baxter integrable
  models in experiments: from condensed matter to ultracold atoms},\
  }\href@noop {} {\bibfield  {journal} {\bibinfo  {journal} {Journal of Physics
  A: Mathematical and Theoretical}\ }\textbf {\bibinfo {volume} {49}},\
  \bibinfo {pages} {173001} (\bibinfo {year} {2016})}\BibitemShut {NoStop}%
\bibitem [{\citenamefont {Gulania}\ \emph {et~al.}()\citenamefont {Gulania},
  \citenamefont {Peng},\ and\ \citenamefont {Govind}}]{gulaniareflection}%
  \BibitemOpen
  \bibfield  {author} {\bibinfo {author} {\bibfnamefont {S.}~\bibnamefont
  {Gulania}}, \bibinfo {author} {\bibfnamefont {B.}~\bibnamefont {Peng}},\ and\
  \bibinfo {author} {\bibfnamefont {N.}~\bibnamefont {Govind}},\ }\bibfield
  {title} {\bibinfo {title} {Reflection symmetry in quantum circuits following
  the {Y}ang-{B}axter equation},\ }\href@noop {} {\bibinfo  {journal} {In
  preparation}\ }\BibitemShut {NoStop}%
\end{thebibliography}%

\end{document}